\documentclass[%
aps,%
prb,%
twocolumn,
showpacs,%
preprintnumbers,%
byrevtex,%
floatfix%
]{revtex4}

\usepackage{epsf}
\usepackage{ifthen}
\usepackage[usenames]{color}
\usepackage{amssymb}
\usepackage{amsfonts}
\usepackage{amsmath}
\usepackage[dvips]{graphicx}

\usepackage[dvips, colorlinks=true, bookmarks, pdftitle={An example PDF file from LaTeX},
 pdfauthor={Gianaurelio Cuniberti}, pdfsubject={From LaTeX to PDF}, pdfkeywords={PDF, LaTeX, hyperlinks, hyperref}]{hyperref}

\def\ii{\textrm{i}\,}

\def\rml{\textrm{L}}
\def\rmr{\textrm{R}}

\def\c{~\cite}

\newcommand{\lab} {\left\langle}
\newcommand{\rab} {\right\rangle}

\begin{document}
\date{April, 18 2006}
\title{Inelastic quantum transport in a ladder model: Measurements of
DNA conduction and comparison to theory}
\author{R.~Guti{\'e}rrez$^{*}$}
\author{S.~Mohapatra$^{*}$}
 \author{H.~Cohen$^{\dagger}$}
  \author{D.~Porath$^{\dagger}$}
\author{G.~Cuniberti$^{*}$}
\affiliation{%
$^{*}$ 
Institute for Theoretical Physics, University of Regensburg, D-93040
Regensburg, Germany \\
$^{\dagger}$ Physical Chemistry Department, The Hebrew University, IL-91904, Jerusalem, Israel
}

\begin{abstract}
We investigate  quantum transport
characteristics of a ladder model, which effectively mimics the
topology of a double-stranded DNA molecule. We consider the interaction of  tunneling charges with a selected internal vibrational degree of freedom and discuss its influence on the structure of the current-voltage characteristics. Further, molecule-electrode contact effects are shown to dramatically affect the orders of magnitude of the current.  Recent electrical
transport measurements on  suspended DNA oligomers with a complex base-pair sequence,  revealing  strikingly high currents, are also presented and used as a reference point for the theoretical modeling. A semi-quantitative description of  the measured $I$-$V$ curves is achieved, suggesting that the coupling to vibrational excitations  plays an important role in DNA conduction. 
\end{abstract}


\maketitle

\section{Introduction}
The past decade has seen an extraordinary progress in the field of molecular
electronics. The possibility  of using single molecules or molecular groups as the basic
building units of electronic circuits has considerably triggered the refinement of
experimental techniques. As a result, transport signatures of individual
molecules have been successfully probed and 
exciting physical effects like
rectification, Coulomb blockade, and the Kondo effect among others have been
demonstrated, see e.g.~Ref.~[\onlinecite{ME05}] for a recent review of the field.

 Within the class of biopolymers, DNA  is expected to play 
 an outstanding role in molecular
 electronics. This is mainly due to its unique self-assembling and
 self-recognition properties, which are essential for its performance as carrier
 of the genetic code, and may be further exploited in the design of electronic
 circuits.~\cite{keren03,pompe99}
 A related important issue is to clarify if DNA in some of its possible
 conformations can carry an electric current or not. In other words, if it could also be
 applied as a wiring system. In the early 1990's 
 charge {\em transfer} experiments
   in natural DNA in solution showed unexpected high charge transfer 
   rates,~\cite{treadway02,murphy93} thus
 suggesting that DNA might support charge transport. However,  {\em electrical transport} experiments carried out on single DNA molecules
  displayed a variety of possible behaviors:
 insulating,~\cite{braun98,storm01} semiconducting,~\cite{porath00,cohen05,cohen06,cohen04} and
 ohmic-like.~\cite{yoo01,tao04} This can apparently be  traced back to the high
 sensitivity of charge propagation in this molecule to extrinsic (interaction with hard substrates, metal-molecule
 contacts, aqueous environment) as well as intrinsic (dynamical structure fluctuations, base-pair sequence)
 factors.
 Recently, experiments on single poly(GC) oligomers in aqueous solution~\cite{tao04} as well as 
 on single suspended  DNA with a complex base sequence~\cite{cohen05,cohen06} have shown unexpectedly 
  high currents of the order of 100-200 nA. These results  strongly suggest that DNA molecules may indeed support rather high electrical currents if the appropriate conditions are warranted. The theoretical interpretation of these experiments and, in a more general context, the mechanism(s) for charge transport in DNA have  not , however,  been revealed so far.

Both  {\em ab initio}
calculations~\cite{artacho03,difelice02,felice04,gervasio02,barnett01,endres05,star04,star05,adessi03,mehrez05}
 as well as model-based  Hamiltonian
 approaches~\cite{gio02,jortner98,jortner02,roche03a,roche03,unge03,hennig04a,jortner05,ac05,ac05a,gmc05a,gmc05b,klh05,yamada04a,bruinsma03}
 have been recently discussed. Though the former can give in principle  a detailed account of
 the electronic and structural properties of DNA, the huge complexity of the molecule
 and the diversity of interactions present in it (internal as well as with the
 counterions and  hydration shells)
 precludes a full systematic first principles treatment of electron transport for realistic molecule  lengths, this becoming even  harder if the dynamic interaction with vibrational degrees of freedom is considered. Thus, Hamiltonian
 approaches can play a complementary role by addressing single factors that  influence
 charge transport in DNA. 
 
 In this paper,  we will address 
 the influence of vibrational
 excitations (vibrons) on the quantum transport signatures of  a ladder model,
 which we use to mimic the  double-strand structure of DNA oligomers. 
 As a reference for our calculations we will take  the previously mentioned  experiments on single 
 suspended DNA molecules with a complex base-pair sequence.~\cite{cohen05,cohen06}
Our main goal is to disclose within a generic Hamiltonian model 
the influence of different parameters on the charge transport properties: the system-electrode coupling, the strength of the 
charge-vibron coupling, and the vibron frequency.  Our model suggests  that strong coupling  to vibrational degrees of freedom may lead  to an enhancement of the zero-current gap, which is  a result of  a vibron blockade effect.~\cite{kvo05}
Further, asymmetries in the ladder-lead coupling have a drastic effect on the absolute values of the current. 
Finally, we show that a two-vibron model can  describe the shape of
the experimental $I$-$V$ curves  of Ref.~[\onlinecite{cohen05}], suggesting that interaction with vibrational degrees of freedom may  give a non-negligible contribution to the measured currents. 
Obviously, other factors related e.g.~to the
 specific metal-molecule interface atomic structure which  govern the efficiency of charge injection, or  the potential profile along the molecule can give  important contributions. 
 They can  be taken into account by a more realistic, 
 fully self-consistent treatment of the
 problem, which lies outside the scope of the present study. 

In the next section we briefly present  the experimental results. In Sec.~{\textbf III.A} the model Hamiltonian is introduced and the relevant parameters are defined. The theoretical formalism is discussed in Sec.~{\textbf  III.B}. Finally, the results are presented and discussed in Sec.~{\textbf  IV}.

\begin{figure}[t]
 \centerline{
 \epsfclipon
 \includegraphics[width=0.99\linewidth]{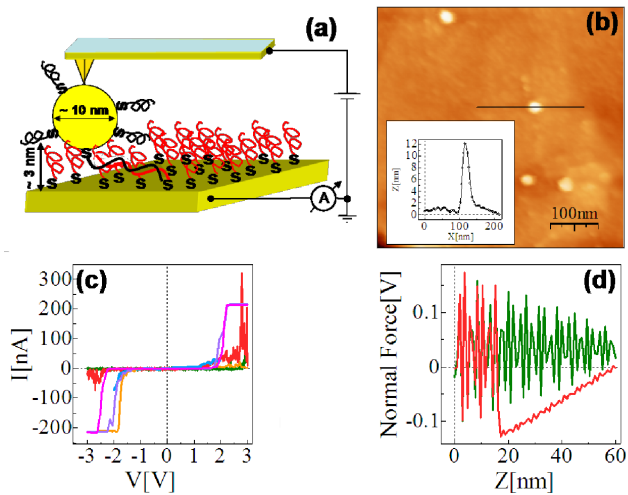}%
 }
 \caption{\label{fig:figCohen}%
 (a) Scheme (not drawn to scale) of the experimental setup showing dithiolated dsDNA (thicker lines for clarity) chemically bonded to two metal electrodes (upper - GNP, lower - gold surface), supported by a monolayer of thiolated ssDNA. (b) AFM topography image showing top view of the sample. Several GNPs are clearly seen on the background of the ssDNA monolayer. The GNPs mark the position of the hybridized dsDNA. The inset is a height profile of the GNP lying on the ssDNA surface. (c) Collection of I-V curves from different samples. In some other cases we measured smaller or no voltage gap. Note that several curves show saturation of the current amplifier at 220 nA. (d) An F-Z curve of one of the curves in (c), (green-forward, red-backward) demonstrating the tip-GNP adhesion (red line) without pressing the GNP through the monolayer. The I-V is recorded at the closest point of the tip to the GNP without pressing it through the ssDNA monolayer.}
 \end{figure}

\section{Experimental aspects}

Detailed description of the sample preparation is reported elsewhere.~\cite{cohen05,cohen06,cohen04} Briefly, a thiolated 26 bases long single-stranded DNA (ssDNA) with a complex sequence was adsorbed on a clean flat annealed gold surface to create a dense monolayer. The ssDNA molecules had a thiol-modified linker end group (CH$_{2}$)$_{3}$-SH at the 3'-end. 
The sequence of the ssDNA that was adsorbed on the gold surface in 0.4 M phosphate buffer with 0.4 M NaCl  is 
5'-CAT\,TAA\,TGC\,TAT\,GCA\,GAA\,AAT\,CTT\,AG-3'-(CH$_{2}$)$_{3}$-SH. 
The surface density of the monolayer on the gold was appropriate for hybridization with complementary thiolated ssDNA that were separately adsorbed on 10 nm gold nanoparticles (GNPs) through another thiol group at their 3'-end by a (CH$_{2}$)$_{3}$-SH group. The monolayer serves also as an insulating support to the GNPs.~\cite{cohen05,cohen06,cohen04} Direct measurements by conductive atomic force microscope (cAFM) with tip-sample bias of up to ±3 V confirmed that the monolayer was insulating. ~\cite{cohen05,cohen06} The double-strand DNA (dsDNA) hybridization was done at ambient conditions in the presence of 25 mM Tris buffer with 0.4 M NaCl.
Before AFM characterization and electrical measurements the samples were thoroughly rinsed to remove excess salts. 

The measurements were done with a commercial AFM (Nanotec Electronica S.L. Madrid) in dynamic mode~\cite{pablo00} to avoid damage to the sample and the metal coated tip. Rectangular cantilevers with Pyramidal tips 
(Olympus, OMCL-RC800PSA, Atomic Force F$\&$E GmbH,  spring constant of 0.3 to 0.7 N/m and resonance frequency of 75 to 80 KHz) were used in order to perform  combined force-distance (F-Z) and current-voltage (I-V) curves with a minimal load on the sample.
For the electrical measurements the tips were sputter coated by Au/Pd that increased their spring constant to about 1 N/m and lowered their resonance frequency to 50-70 KHz. Throughout the measurements the cantilever was oscillated close to its resonance frequency and feedback was performed on the amplitude of its vibrations.
Fig.~1(a) shows a schematic view of the sample and set-up. Fig. 1(b) is an AFM image showing several GNPs, indicating the position of the hybridized dsDNA on the background of the ssDNA monolayer. A line profile along one of the  10 nm GNPs implies that the 26 bp dsDNA, which is $\sim$9 nm long, is not protruding vertically out of the $\sim 3-4$ nm thick ssDNA monolayer and is probably tilted and lying on the surface of the ssDNA monolayer.
The electrical I-V curves (Fig.~1(c)) were recorded while the GNP was contacted during an F-Z curve by the metal covered tip without pressing the tip onto the GNP. This was done by applying a feedback on the tip oscillation amplitudes, while approaching the GNP, that enabled to stop the tip movement towards the GNP just before the jump to contact, as demonstrated in the F-Z curve showed in Fig. 1(d).

The current voltage curves, shown in Fig 1(c), demonstrate in a clear and reproducible way, the ability of $\sim$9 nm long dsDNA to conduct relatively high currents ($>$ 200 nA), when the molecule is not attached to a hard surface along its backbone and when charge can be injected efficiently through a chemical bond. Such behavior was measured for many dsDNA molecules on tens of samples and with various tips and humidity conditions, with similar results.~\cite{cohen05} This behavior was also measured in the absence of the GNPs using a different technique.~\cite{cohen06}


\section{Theoretical aspects}

\subsection{Model Hamiltonian}
\begin{figure}[t]
\centerline{
\epsfclipon
\includegraphics[width=.99\linewidth]{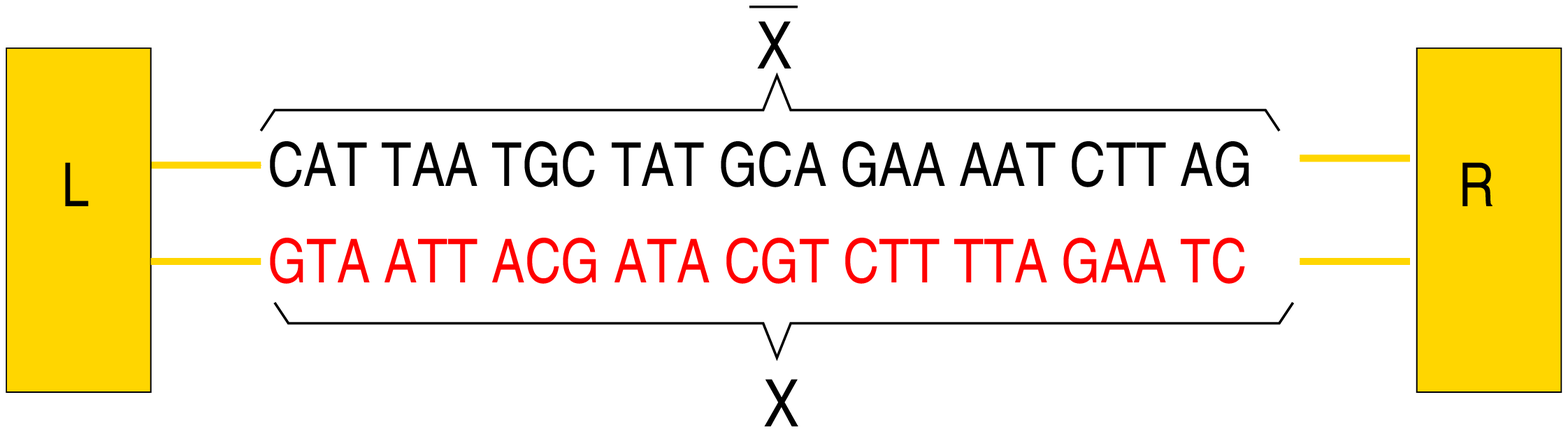}%
}
\centerline{
\epsfclipon
\includegraphics[width=.99\linewidth]{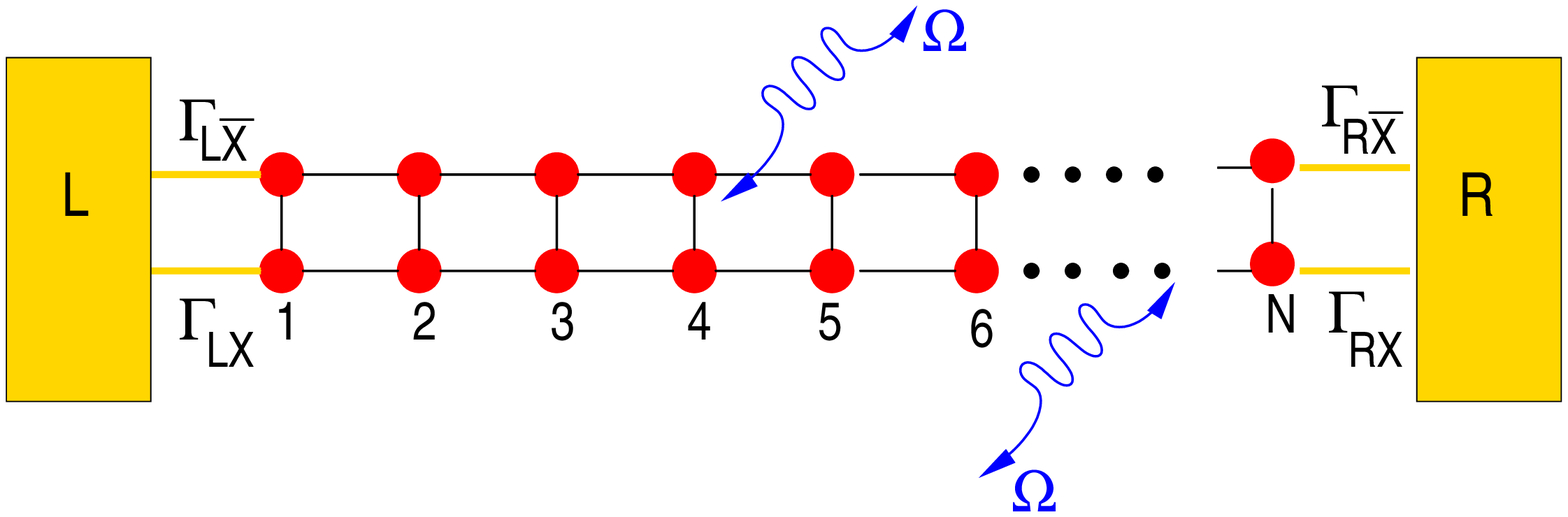}%
}
\caption{\label{fig:fig2}%
Upper panel: Schematic representation of the double-strand  DNA with the experimentally relevant base-pair sequence.~\cite{cohen05} The (CH$_{2}$)$_{3}$-SH linker groups are omitted for simplicity (see the text for details).  Lower panel: Two-legs ladder  used to mimic the double-strand
structure of a DNA molecule. L and R refer to left and right electrodes, respectively. The coupling terms to the electrodes $\Gamma_{\ell,\alpha},\ell=$X,$\bar{X}$, $\alpha=$L,R are assumed to be energy-independent constants (wide -band limit, see the text for details).
}
\end{figure}

Our aim is to formulate a {\em minimal} model  taking into account the double-strand structure of DNA.
 Hence, we do not consider the full complexity of the 
DNA electronic structure. We  neglect environmental effects  and  assume that charge transport will mainly take place along the base-pair stack. We further adopt the  perspective that to describe 
low-energy quantum transport within a single-particle picture, only the frontier $\pi$ orbitals of the base pairs
are relevant. We  will then  consider a planar 
ladder
model with a single orbital per lattice site within a nearest-neighbor tight-binding picture. 
In this sense, we are
neglecting helical effects arising from the real structure of the DNA. We assume that these and similar effects  may have already renormalized the electronic  parameters.
We will   focus in this paper on the experimentally relevant\cite{cohen05}  base-pair sequence  
X=5'-CAT\,TAA\,TGC\,TAT\,GCA\,GAA\,AAT\,CTT\,AG-3', see Fig.~\ref{fig:fig2} and the foregoing section.
It is worth mentioning  that ladder models have been previously used 
to study quantum transport in DNA
duplexes.~\cite{yi03,yamada04,yamada04a,klotsa05,caetano04,tapash06}

The Hamilton operator describing the ladder and its coupling to left (L) and right (R)
electronic reservoirs is given by: 

 \begin{eqnarray}
\cal{H}_{\rm el}&=& 
\sum_{\rm{r}=X,\bar{X}}\sum_{\ell} \epsilon_{\rm{r},\ell} 
b^{\dagger}_{\rm{r},\ell} b_{\rm{r},\ell} \nonumber \\
&-&\sum_{\rm{r}=X,\bar{X}}\sum_{\ell} t_{\rm{r},\ell,\ell+1} [ b^{\dagger}_{\rm{r},\ell}
b_{\rm{r},\ell+1} + \textrm{h.c.} ] \nonumber \\
&-&\sum_{\ell} t_{\perp,\ell} [ b^{\dagger}_{\rm{X},\ell}
b_{\rm{\bar{X}},\ell} +  \textrm{h.c.} ]\nonumber \\
&+&\sum_{{\bf k}\in L} [ t_{{\bf k},X} c^{\dagger}_{{\bf k}} b_{X,1} +  \textrm{h.c.}]
+\sum_{{\bf k}\in L} [ t_{{\bf k},\bar{X}} c^{\dagger}_{{\bf k}} b_{\bar{X},1} + \textrm{h.c.} ]\nonumber \\
&+&\sum_{{\bf k}\in R} [ t_{{\bf k},X} c^{\dagger}_{{\bf k}} b_{X,N} +  \textrm{h.c.}]
+\sum_{{\bf k}\in R} [ t_{{\bf k},\bar{X}} c^{\dagger}_{{\bf k}} b_{\bar{X},N} + \textrm{h.c.}
]\nonumber
\end{eqnarray}
 In the previous expression, $X,\bar{X}$ refer to the two legs of the ladder,
 $\epsilon_{\rm{r},\ell}$ are  energies at site $\ell$ on leg $\rm{r}$, 
 $t_{\rm{r},\ell,\ell+1}$ are the corresponding nearest-neighbor electronic hopping integrals along the two
 strands while  $t_{\perp,\ell}$ describes the inter-strand hopping. 
In order to obtain  estimates   of onsite energies and hopping integrals, {\em
ab initio } calculations are obviously the most reliable reference point. Recently, Mehrez and
Anantram~\cite{mehrez05} carried out a careful analysis of a hierarchy of
tight-binding models that gave effective onsite energies and hopping parameters for
Poly(GC) and Poly(AT) molecules. We
will use these values as a reference point in part of our discussion and take the onsite energies 
as the LUMO energies given in Ref.~[\onlinecite{mehrez05}]: $\epsilon_{\rm G}=1.14 \,$eV,$\epsilon_{\rm C}=-1.06 \,$eV, $\epsilon_{\rm A}=0.26 \, $eV,$\epsilon_{\rm T}=-0.93 \,$eV. We are thus considering electron transport, although hole transport can be dealt with in a similar way by choosing the HOMO instead of the LUMO energies. Other choices e.~g.~the ionization potentials of the base pairs are also possible;\c{roche03} they are expected to  only change our results quantitatively. More difficult is the choice of the  intra- and inter-strand electronic transfer integrals. They will be more sensitive to the specific base sequence considered.   For the sake of simplicity and in order to reduce the number of model parameters we have adopted a  simple  parameterization taking a  homogeneous hopping along both legs, i.~e.~$t_{\rm{r},\ell,\ell+1}=t_{\rm{X}}=t_{\rm{\bar{X}}}=t\sim 0.25-0.27$ eV and $t_{\perp,\ell}=t_{X\bar{X}}\sim$0.2-0.3 eV. 
 Though calculations~\cite{mehrez05,bixon00} show that the inter-strand hopping is usually very small, $\sim$ few \textrm{meV}, we do not consider the hopping integrals  as  bare tight-binding parameters but as  effective ones, thus keeping some freedom in the choice of their specific values. Electronic correlations~\cite{yi03} or structural fluctuations mediated by the coupling to other vibrational degrees of freedom~\cite{voityuk01} can lead to a strong  renormalization of the bare electronic coupling.

 The interaction with the electronic reservoirs will be described in the most simple
 way by invoking the wide-band approximation, i.~e.~neglecting the energy dependence
 of the leads' self-energies (see below).
 To model the coupling to vibrational degrees of freedom we consider the case 
 of  long-wave length  optical modes with constant frequencies $\Omega_{\alpha}$, e.~g.  small-{\bf q} torsional modes and assume  they  couple to the total charge density operator $N=\sum_{\rm r, \ell}n_{\rm r,\ell}$ of the ladder. This approximation
 can be justified for long-wave length distortions. In other words, the strength of
 the electron-vibron interaction $\lambda$ is assumed to be site-independent. Moreover, we will not consider in this study non-local coupling to vibrational excitations. Though this interaction can give an important contribution to the modulation of the inter-site electronic hopping, its inclusion would increase the complexity of the model and the number of free parameters. Such effects deserve a separate investigation; research along these lines has been recently presented by other authors.~\cite{berlin02,berlin01,grozema00,hennig04,siebbeles05}
The total Hamiltonian thus reads:
 
\begin{eqnarray}
{\cal{H}}= {\cal{H}_{\rm el}}+
\sum_{\alpha} \Omega_{\alpha} B^{\dagger}_{\alpha} B_{\alpha}
+\sum_{\rm{r},\ell,\alpha }\lambda_{\alpha} 
 b^{\dagger}_{\rm{r},\ell} b_{\rm{r},\ell} (B_{\alpha}+B^{\dagger}_{\alpha})
 \label{eq:eq1}
\end{eqnarray}

\subsection{Green function techniques}

In this section we present the theoretical approach to deal with  electrical transport 
properties in the presence of electron-vibron coupling. 
Taking as a starting point the Hamiltonian of Eq.~\ref{eq:eq1}, we perform 
a Lang-Firsov (LF) unitary transformation~\cite{mahan} in order to eliminate the electron-vibron interaction. 
The LF-generator is given by 
${\cal U}=\exp[-\sum_{\alpha,\rm{r},\ell}
g_{\alpha} b^{\dagger}_{\rm{r},\ell} b_{\rm{r},\ell} (B_{\alpha}-B^{\dagger}_{\alpha})]$, which is basically a shift operator for the harmonic oscillator position. The parameter $g_{\alpha}=\lambda_{\alpha}/\Omega_{\alpha}$ gives an effective measure of the electron-vibron coupling strength. 
In the
resulting Hamiltonian, the onsite energies  $\epsilon_{\rm{r},\ell}$ are shifted
to $\epsilon_{\rm{r},\ell}-\Delta$ with
 $\Delta=\sum_{\alpha}\lambda^2_{\alpha}/\Omega_{\alpha}$ 
being the polaron shift. 
 There is an
 additional renormalization of the tunneling Hamiltonian, but we will not consider
 it explicitly, since we will  assume a  regime (within the  wide-band approximation 
 in the leads' spectral densities) where the effective broadening
 $\sim\Gamma$ arising from the coupling to the leads is  bigger than
 the polaron formation energy $\sim\lambda^2/\Omega$ . As shown in
 Ref.~[\onlinecite{hewson79}]  in this special case the tunneling renormalization can be approximately neglected.
 
Concerning the transport problem, we can use the standard current expression for lead $p$=L,R  as derived e.g.~by Meir and Wingreen.~\cite{mw92}
\begin{eqnarray}
 I_p = \frac{2\ii e}{h}\int dE \,\textrm{Tr} [\Gamma_{\textrm p}\{ f_{\textrm p}
 (G^{>}-G^{<})+ G^{<}\} ],
 \label{eq:eq2}
 \end{eqnarray}  
and then perform the LF unitary transformation under the trace going over to the transformed Green functions.  In the previous equation,  $\Gamma_{\textrm p}(E)=\ii \,(\Sigma_{\textrm p}(E)-\Sigma^{\dagger}_{\textrm p}(E))$ are the leads'
 spectral functions, $f_{\textrm p}(E)=f(E-\mu_{\textrm p})$ is the Fermi function of the
 $\textrm{p}$-lead and $\mu_{\textrm {p}=\rml}=E_{\rm F}+ eV/2 \;(\mu_{\textrm {p}=\rmr}=E_{\rm F}-eV/2)$ are
  the
 corresponding electrochemical potentials. We assume hereby a symmetrically applied bias. 
Within the wide-band limit in the electrodes' spectral densities, we introduce the following  $2N\times 2N$ ladder-lead energy-independent coupling matrices: \\
\[   (\Gamma_{\textrm{L}})_{nm}=\left \{ 
\begin{array} 
{r@ { \quad \textrm{if}  \quad }l}
\Gamma_{L,X} \delta_{n,1}\delta_{m,1} & n,m \in X \\
\Gamma_{L,\bar{X}} \delta_{n,1}\delta_{m,1} & n,m \in \bar{X} \\
0 & \neq 1
\end{array}
\right. \]

\[  (\Gamma_{\textrm{R}})_{nm}=\left \{ 
  \begin{array} 
  {r@ { \quad \textrm{if}  \quad }l}
  \Gamma_{R,X} \delta_{n,N}\delta_{m,N} &  n,m \in X \\
  \Gamma_{R,\bar{X}} \delta_{n,N}\delta_{m,N} & n,m \in \bar{X} \\
  0 & n,m \neq N
  \end{array} 
  \right. \]

We remark  at this point that these coupling terms also include effectively the (CH$_{2}$)$_{3}$-SH linkers used in the experiments to attach the DNA molecule to the metallic electrodes.. 

 Let's define the fermionic vector operator (see Fig.~\ref{fig:fig2} for reference):
 \begin{eqnarray}
 \Psi^{\dagger}=(b_{X,1}\, b_{X,2} \,\cdots b_{X,N}\, b_{\bar{X},1} \, \cdots \,b_{\bar{X},N} ).
 \end{eqnarray}  
 The lesser- and
 greater-matrix Green function (GF) are then  defined as:
  \begin{eqnarray}
  G^{>}(t)&=&-\frac{\ii}{\hbar} \lab \Psi(t) \Psi^{\dagger}(0) \rab,  \\
  G^{<}(t)&=& \frac{\ii}{\hbar} \lab \Psi^{\dagger}(0) \Psi(t) \rab. \nonumber
  \end{eqnarray}   
 
 Since
 Eq.~\ref{eq:eq2} does not explicitly contain information on the specific
 structure of the ``molecular'' Hamiltonian, we can now transform the lesser- and
 greater-GF as well as the lead spectral functions  to the polaron representation. 
 The operator $\Psi$ transforms according to $\bar{\Psi}={\cal U}\Psi {\cal
 U}^{\dagger}=\Psi {\cal X}$, where ${\cal X}=\exp[\sum_{\alpha}
(\lambda_{\alpha}/\Omega_{\alpha})(B_{\alpha}-B^{\dagger}_{\alpha})]$. Thus, we
obtain  $\bar{G}^{>}(t)=-(\ii/\hbar) \lab \Psi(t){\cal X}(t) \, \Psi^{\dagger}(0){\cal X}^{\dagger}(0)
\rab$ and similar for $\bar{G}^{<}(t)$. Strictly speaking, a further direct decoupling of the
foregoing expression into purely fermionic and vibronic components, as is usual in the independent vibron model~\cite{mahan} is not
possible, since the transformed 
tunneling Hamiltonian contains both types of operators and hence, the
transformed canonical density operator does not factorize into separate fermion
and vibron density operators. However, for the case considered here,  
where  vibron-induced renormalization
effects of the tunneling amplitudes are not taken into account, 
the decoupling is  still approximately possible. We thus obtain: 

 \begin{eqnarray}
 \bar{G}^{>}(t)&=&-\frac{\ii}{\hbar} \lab \Psi(t){\cal X}(t)\, \Psi^{\dagger}(0){\cal X}^{\dagger}(0)\rab \nonumber  \\
&\approx& -\frac{\ii}{\hbar}\lab \Psi(t)\Psi^{\dagger}(0)\rab_{el} \lab {\cal X}(t) {\cal X}^{\dagger}(0)\rab_{B}
 \nonumber \\
 &=&
 G^{>}(t)\lab {\cal X}(t) {\cal X}^{\dagger}(0)\rab_{B}=G^{>}(t) e^{-\Phi(t)}. \nonumber 
  \end{eqnarray}
     
A similar expression holds for the lesser-than GF by changing the time argument 
$t$ by $-t$ in
$\Phi(t)$. We note that $\Phi(t)$ satisfies the symmetry relations:
$\Phi(-t)=\Phi^{\dagger}(t)$.

\subsection{Single-vibron case}

In  the case of  dispersionless modes, 
 the vibron correlation function $\Phi(t)$ can be evaluated exactly and reads:~\cite{mahan}

 \begin{eqnarray}
 e^{-\Phi(t)}=e^{-g^2(2N+1)} \, \sum^{\infty}_{n=-\infty} I_{n}(\tau) e^{\beta\Omega n/2}
e^{-\ii n\Omega t},
 \end{eqnarray}
 where $ \tau= 2g^2 \sqrt{N(N+1)}$ and $g=\lambda/\Omega$. 
 Using this expression, one easily finds for the Fourier transformed lesser and greater GFs:
  \begin{eqnarray}
 \bar{G}^{<(>)}(E)&=&\sum^{\infty}_{n=-\infty} \phi_{n}(\tau) G^{<(>)}(E+(-)n\Omega), \\
 \phi_{n}(\tau)&=&e^{{-g^{2}(2N+1)}} \;\times I_{n}(\tau)
\; e^{{\beta\Omega\,n/2}}. \nonumber \\
\label{eq:tau}
   \end{eqnarray}
 where $+(-)$ corresponds to $<(>)$. 
 The bare lesser- and greater-GF can now be obtained from  the kinetic equation 
 $G^{<(>)}=G^{r}(\Sigma^{<(>)}_{\rml}+\Sigma^{<(>)}_{\rmr})G^{a}$, since the full 
 electron-vibron
 coupling is already contained in the  prefactor function $\phi_n(\tau)$. The leads self-energy matrices $\Sigma^{<}_{\textrm p},\Sigma^{>}_{\textrm p}$ are given in the wide-band limit by $\ii f_{\textrm{p}}(E)\Gamma_{\textrm p}$  and $-\ii (1-f_{\textrm{p}}(E))\Gamma_{\textrm p}$, respectively.
 Using these expressions, the total symmetrized current
 in the stationary state $J_{\rm T}=(J_{\rml}-J_{\rmr})/2$ is given by (see
 Appendix): 
 
  \begin{eqnarray} 
J_{\rm T}&=&\frac{e}{2h} \sum^{\infty}_{n=-\infty} \phi_{n}(\tau) 
   \int dE \, \{ [f_{\rml}(E) \, (1-f_{\rmr}(E-n\Omega)) \nonumber \\ &-&
   f_{\rmr}(E) \, (1-f_{\rml}(E-n\Omega)) ]\, t(E-n \Omega) \nonumber\\
  &+& [f_{\rml}(E+n\Omega) \, (1-f_{\rmr}(E)) \nonumber \\ &-& 
   f_{\rmr}(E+n\Omega)\, (1-f_{\rml}(E)) ]\, t(E+n \Omega)  \}, 
 \label{eq:eqcur}  
 \end{eqnarray}

where $t(z)= \textrm{Tr}[\Gamma_{\rmr} G^{\textrm{r}}(z) \Gamma_{\rml} G^{\textrm{a}}(z)]$ is the
conventional expression for the transmission coefficient in terms of the 
molecular  Green function $G(E)$, which satisfies the Dyson-equation: $G^{-1}=G^{-1}_{0}-\Sigma_{\textrm{L}}-\Sigma_{\textrm{R}}$. 
The above result for the current has a clear physical
 interpretation. So, e.g.~a term like
 $f_{\rml}(E) \, (1-f_{\rmr}(E-n\Omega))t(E-n \Omega)$ describes an electron 
 in the left lead which tunnels into the molecular region, emits $n$ vibrons of
 frequency $\Omega$ and tunnels out into the right lead. However, it can only go into
 empty states, hence the Pauli blocking factor $(1-f_{\rmr}(E-n\Omega))$. Other
 terms can be interpreted along the same lines, when one {\it additionally} 
 substitutes electrons by holes. 

Finally, the spectral density $A(E,V)$ is defined as:
\begin{eqnarray}
A(E,V)&=&\ii \, [\bar {G}^{>}(E)-\bar{G}^{<}(E)]\\
&=&\ii\, \sum_{n}\phi_n(\tau)\,  [G^{>}(E-n\Omega)-{G}^{<}(E+n\Omega)]   \nonumber
\end{eqnarray}

\section{Results}
\begin{figure}[t]
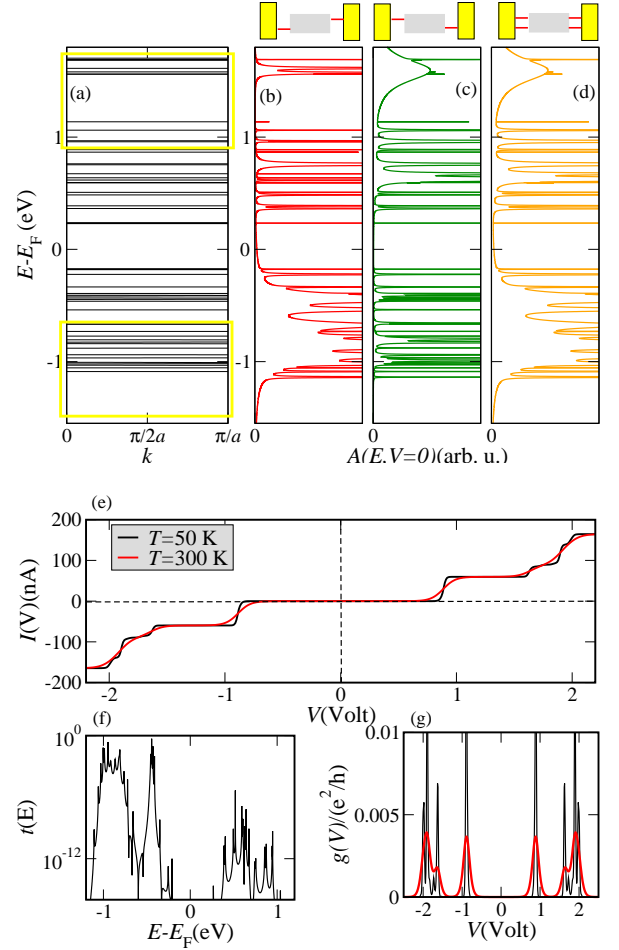

\centerline{
\epsfclipon
\includegraphics[width=.90\linewidth]{./Fig_3a-d.eps}%
}\vspace{0.3cm}
\centerline{
\epsfclipon
\includegraphics[width=.90\linewidth]{./Fig_3e-g.eps}%
}
\caption{\label{fig:fig3}%
(a) Tight-binding electronic band structure of
an infinite DNA system, obtained by a periodic repetition of the 26-base
sequence of Ref.~[\onlinecite{cohen05}]. Notice the strongly fragmented band structure with very
flat bands. The open yellow rectangles indicate for reference 
the approximate position of the bands for a periodic poly(GC) oligomer. 
(b)-(d)  spectral density $A(E,V=0)$, which at zero voltage coincides with the projected density of states onto the molecular region, for  the {\it finite size} DNA chain  contacted in different ways by left and right electrodes, see Fig.~\ref{fig:fig2}: (b)$\Gamma_{L,X}=\Gamma_{R,\bar{X}}=0,\Gamma_{L,\bar{X}}=\Gamma_{R,X}=250\,$meV, 
(c) $\Gamma_{L,X}=\Gamma_{R,\bar{X}}=250\,$meV,$\Gamma_{L,\bar{X}}=\Gamma_{R,X}=0\,$meV, 
(d) $\Gamma_{L,X}=\Gamma_{R,\bar{X}}=\Gamma_{L,\bar{X}}=\Gamma_{R,X}=250\,$meV.
The onsite energies were set at the LUMO values reported in
Ref.~[\onlinecite{mehrez05}] and the hopping parameters were set to $t_{X}=t_{\bar{X}}=t=0.27\,$eV, $t_{X\bar{X}}=0.25\,$eV. 
(e)  $I$-$V$ characteristics for two different temperatures and the contact situation (b); (f) corresponding transmission function $t(E)$ and  (g) differential conductance $g(V)$.}
\end{figure}

In Fig.~\ref{fig:fig3}(a) 
we first show the electronic band structure of an {\it infinite} periodic array of the 
26-base-pairs DNA molecule without considering charge-vibron interactions. The
unit cell thus contains  2$\times$26 sites. Due to the large unit cell and since the electronic hopping integrals are roughly a factor four  smaller than the onsite 
energies,  one gets a strongly fragmented electronic spectrum with very flat bands. 
We may thus rather speak of
valence and conduction manifolds as of true dense electronic
bands.~\cite{felice02} The band gap $\Delta$ of about 0.3 eV is considerably
smaller than that obtained in the periodic poly(GC) ladder when using  the {\em same} parameterization, 
$\Delta_{\rm{GC}}\sim 2.0$ eV. In Fig.~\ref{fig:fig3}(a) we also show schematically the 
positions of the conduction and valence manifolds of the periodic poly(GC) system (open
rectangles). We note in passing that similar small gaps have been estimated in experiments on $\lambda$-DNA~\cite{tran00} and in bundles~\cite{rakitin01}. A direct comparison to our results is however not possible due to the different experimental conditions and length scales probed in these investigations. 
Figs.~\ref{fig:fig3}(b)-(d) show the spectral density at zero voltage of the  {\it finite} DNA ladder contacted by  electrodes in three different ways:
(b) only the 3'-ends, (c) only the 5'-ends, and (d) all four ends of the double-strand are contacted. 
Though the general effect consists in broadening of the electronic manifolds, we also see that depending on the way the molecule is contacted to the leads the electronic states will be affected in different ways. Thus e.g.~states around 1.7 eV above the Fermi level are considerably  more broadened than states closer to E$_{\textrm{F}}$.  Figs.~\ref{fig:fig3}(e)-(g)  show the current, transmission and differential conductance for one of the typical contact situations (case (b)).  The irregular step-like structure in the current-voltage characteristics is reflecting the fragmented electronic structure of the 
system.  Notice that despite the small gap found in the band structure resp. DOS, a large ($\sim
2$ V) zero-current gap is seen in the $I$-$V$ characteristics. The reason is that many of the
states close to the band gap have a very low transmission probability (are highly localized) as a result of the random base sequence, see $t(E)$ in  Fig.~\ref{fig:fig3}(f), so that they do not contribute to transport. The   effect of the temperature is only to smooth the current and the differential conductance, as expected. We remark at this point  that the absolute values of the current can be  dramatically modified by the way the molecule is contacted by the electrodes (see below).

We now consider the coupling to 
vibrational degrees of freedom in the ladder. The probability of opening inelastic transport channels
by emission or absorption of $n$ vibrons becomes higher 
with increasing thermal energy  $k_{\textrm B}T$ and/or
electron-vibron coupling $g$. As a result, the spectral
density $A(E)$ will consist of a series of elastic peaks (corresponding to $n=0$) 
plus
vibron
satellites ($n\neq 0$). If the separation between contiguous
molecular eigenstates is of the  order of the vibron frequency
$\Omega$, then the satellites corresponding to a given molecular state will not be clearly separated from 
the elastic peaks   but will overlap
with those of nearby molecular states leading to an effective broadening of the spectrum and possibly to complex interference effects.

\begin{figure}[t]
\centerline{
\includegraphics[width=.99\linewidth]{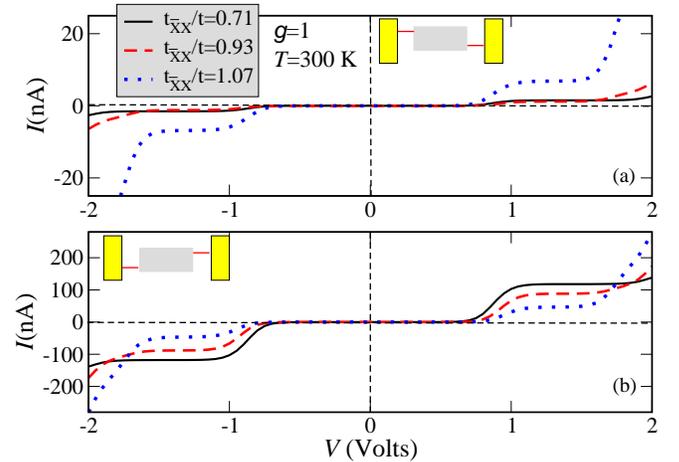}%
}
\caption{\label{fig:fig4}
$I$-$V$ characteristics for different   values of the reduced inter-strand hopping $t_{X\bar{X}}/t$  in
DNA$_{26}$ for a fixed electron-vibron coupling strength $g=1$. Upper and lower panels correspond to two different (asymmetric) ways of coupling the ladder to the electrodes: 
(a) $\Gamma_{L,X}=\Gamma_{R,\bar{X}}=0, \Gamma_{L,\bar{X}}=\Gamma_{R,X}=250\,$meV and 
(b)$\Gamma_{L,\bar{X}}=\Gamma_{R,X}=0, \Gamma_{L,X}=\Gamma_{R,\bar{X}}=250\,$meV.  
Notice the strong variation  in the current when going from case (a) to case (b). }
\end{figure}

The influence of the transverse electronic hopping and the leads-ladder coupling on the current is shown in Fig~\ref{fig:fig4}. The inter-strand hopping  turns out to be  
crucial in determining the absolute values of the current. In the general case of symmetric coupling ($\Gamma_{L,X}=\Gamma_{R,X}=\Gamma_{L,\bar{X}}=\Gamma_{R,\bar{X}}$),  a charge propagating along one of the strands 
will see a rather disordered system, so that a non-zero  inter-strand hopping may increase the delocalization of  the electronic states. The effect should be more obvious in the  case of asymmetric coupling (a) $\Gamma_{L,\bar{X}}=\Gamma_{R,X}\neq 0, \Gamma_{L,X}=\Gamma_{R,\bar{X}}=0$, and (b) $\Gamma_{L,\bar{X}}=\Gamma_{R,X}= 0, \Gamma_{L,X}=\Gamma_{R,\bar{X}}\neq 0$ since now there is a  single pathway for an electron tunneling from the electrodes into the ladder, e.g.~$L\to X\to  \bar{X}\to R$. We thus see in 
Fig.~\ref{fig:fig4} that relative small variations  of $t_{X\bar{X}}$ considerably modify the current. We note in passing that recent transport measurements on DNA oligonucleotides  have displayed considerable differences in the conductance of single- vs. double-stranded DNA, thus suggesting  that, apart from other factors, inter-strand interactions may  play a role in controlling charge transport.~\cite{zalinge06} 

Our above results are, moreover, very sensitive to the way the ladder is coupled to the electrodes, as seen from the upper and lower panels of Fig.~\ref{fig:fig4}.
These cases  are related respectively to the situation where only 
the 5'-end  (a) or only  the 3'-end  sites (b) of the ladder have
non-zero coupling to the electrodes, see Fig.~\ref{fig:fig2} for reference.  Notice
that  (b) would correspond to the  experimental contact geometry  in
Ref.~[\onlinecite{cohen05}] where  only the 3'-end of each strand in the double helix is connected $-$ via the linker groups $-$ to one of the electrodes (Au-substrate and GNP).  
Configuration (b) also leads to  considerably higher currents  than the case (a).

More generic assertions require, however, a detailed atomistic investigation of the DNA-metal contact topology and base-pairs energetics, which  goes beyond the scope of this study.

\begin{figure}[t]
\centerline{
\epsfclipon
\includegraphics[width=.99\linewidth]{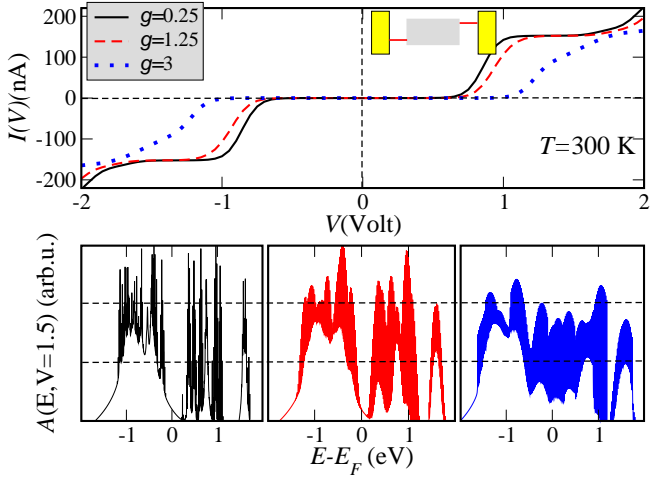}%
}
\caption{\label{fig:fig5}%
Dependence of the current on the effective electron-vibron coupling strength $g=\lambda/\Omega$ at $T=300$ K and for $t_{X\bar{X}}=0.71\,t$. The vibron frequency was fixed at 20 meV.
With increasing coupling the total current is reduced and the zero-current gap is enhanced. 
The lower panels show the spectral density at  $V\sim 1.5\,$Volt for the three values of $g$. Despite the increased 
number of vibron satellites with increasing coupling, the total intensity is reduced. }
\end{figure}

Fig.~\ref{fig:fig5} shows the influence of the coupling to the vibron mode on the magnitude of the current and of 
the zero-current gap.   The slope of the $I$-$V$ curves
is considerably  reduced with increasing $g$.  The corresponding spectral densities at $V\sim 1.5$ Volt, see  Fig.~\ref{fig:fig5}, lower panel, show  broadening due to the emergence of  an increasing number of  vibron satellites (inelastic channels)  with larger coupling, but at the same time a 
redistribution of spectral weights takes place. This is simply the result of the sum rule $\int dE\,A(E)=2\pi$.  
The reason for the current reduction can be qualitatively understood by looking at the spectral density. The reduction in the intensity of $A(E)$ will clearly lead to a reduction in the current at a fixed voltage, since it is basically the area under $A(E,V=\textrm{const.})$ within the energy window $\left[E_{\textrm{F}}-eV/2, E_{\textrm{F}}+eV/2\right]$  which really matters.  
Notice also the increase of the zero-current gap with increasing electron-vibron coupling (vibron blockade), which is related to the exponential suppression of transitions between low-energy vibronic states.~\cite{kvo05}Alternatively, this can be interpreted as an increase of the effective mass of the polaron which thus leads to its localization and to a blocking  of transport at low energies. 

\begin{figure}[t]
\centerline{
\epsfclipon
\includegraphics[width=.99\linewidth]{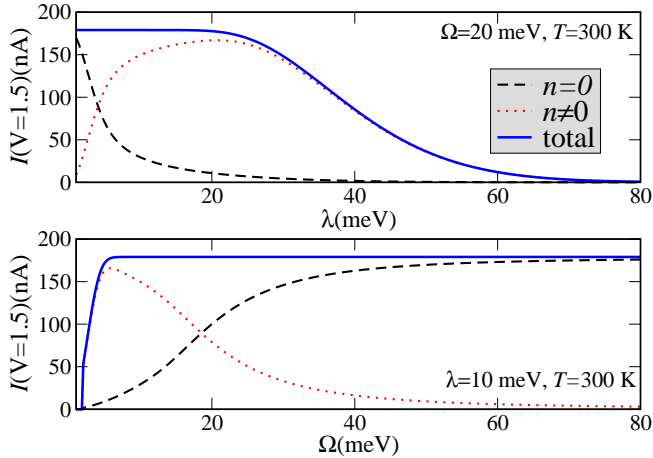}%
}
\caption{\label{fig:fig6}%
Dependence of the elastic ($n=0$) and inelastic ($n\neq 0$) components of the current at a {\it fixed} voltage  on the electron-vibron coupling strength $\lambda$ and the mode frequency  $\Omega$. A more detailed 
analysis of the behavior is presented in the text. The dashed lines correspond to the total current (sum of elastic and inelastic components)}
\end{figure}

Fig.~\ref{fig:fig6} shows in a more systematic way the influence of $\lambda$ and $\Omega$ on the elastic and inelastic components of the total current.  
The dependence on $\lambda$ is easier to understand since only the prefactors $\phi_{n}(\tau)$ do depend on it. One can  show that 
$\phi_{n=0}(\tau)$ is a monotonous decreasing function of  $g$(or $\lambda$), while  $\phi_{n\neq 0}(\tau)$ grows up first, reaches a maximum,  and then exponentially decays for larger $g$. As a result, the elastic  current starts at its bare value for zero coupling to the vibron mode and then it  rapidly decreases when increasing the coupling, because the probability for emission/absorption of vibrons accordingly increases. The   inelastic component, on the other hand, will first increase for moderate coupling and thus gives the dominant contribution to the total current over some intermediate range of $g$'s (which will also depend on the temperature and the vibron frequency). For even larger $g$'s the inelastic current also goes to zero and the current 
will be finally  suppressed, since  there is  an increasing  trend to charge localization with increasing coupling to the vibron.  The behavior at large frequencies is also  plausible, see lower panel of  Fig.~\ref{fig:fig6}, since the average distance between the elastic peak and the inelastic channels is of the order $n\Omega$; if $\Omega$ is large enough  an electron injected with a given energy (fixed voltage) will not be able to excite vibrons in the molecular region and thus only the elastic channel will be available. Alternatively, a very stiff mode ($\Omega\to\infty$) will clearly have no influence on the transport. For very low $\Omega$ the inelastic current will obviously vanish, but the elastic component should simply go over into its bare value without charge-vibron coupling. The fact that also the elastic part goes to zero in Fig.~\ref{fig:fig6}, lower panel, is simply an artifact related to the fact that  at  $\Omega=0$ the LF transformation is ill-defined.  Since we  only consider finite frequencies, this limiting case is not relevant for our discussion.Technical details  are presented in Appendix {\bf B}.

We finally show that extending the previous model to include two
vibrational excitations allows for a semi-quantitative description of
the  experimental results of Ref.~[\onlinecite{cohen05}]. One should, however, keep in mind that these calculations are not giving an explanation of the high currents observed; from the experimental point of view there are some  issues like the number of molecules contacted or the specific details of the DNA-metal contacts  which are not completely clarified. Our aim is rather to point out  at the possible influence  of vibrational degrees of freedom in these recent experiments.  
Using the formalism of Sec.~{\bf II} it  is straightforward to obtain expressions for the current in the two-vibron  case. One finds ($g_{\rm{s}}= \lambda_{\rm{s}}/\Omega_{s}$, s=1,2): 

\begin{eqnarray}
j_{\textrm{tot}}(V)&=&\frac{e}{2h} \sum^{\infty}_{n=-\infty} \sum^{\infty}_{m=-\infty}
\phi_{n,1}\;\phi_{m,2}   \int \, dE \,  \times \nonumber \\ 
&& \{ [  f_{\textrm{L}}(E)(1-f_{\textrm{R}}(E-(n\Omega_1+m\Omega_2)))\nonumber \\ 
&-& f_{\textrm{R}}(E) (1-f_{\textrm{L}}(E-(n\Omega_1+m\Omega_2))) ] \nonumber \\ &\times& t(E-(n\Omega_1+m\Omega_2))  \nonumber \\
 &+&  [f_{\textrm{L}}(E+(n\Omega_1+m\Omega_2))(1-f_{\textrm{R}}(E)) \nonumber \\
& -& f_{\textrm{R}}(E+(n\Omega_1+m\Omega_2))(1-f_{\textrm{L}}(E)) ]\nonumber \\ &\times& t(E+(n\Omega_1+m\Omega_2)) \}  \nonumber \\
\phi_{n,\rm{s}}(\tau)&=&e^{{-g^{2}_{\rm{s}}(2N_{s}+1)}} \;\times I_{n}(\tau_{s})
\; e^{{\beta\Omega_{s}\,n/2}}. \nonumber 
\end{eqnarray}

\begin{figure}[t]
\includegraphics[width=.95\linewidth,angle=0]{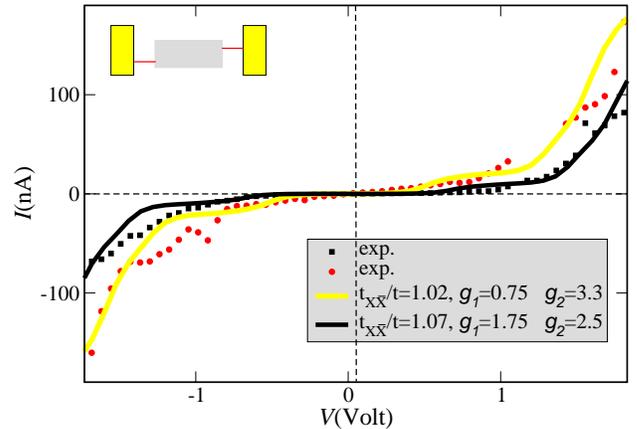}%
\caption{\label{fig:fig7}%
Theoretical curves (solid lines) compared with two different $I$-$V$ curves as obtained on suspended double-strand DNA oligomers contacted by a GNP.~\cite{cohen05} In both cases the temperature and the coupling to the electrodes were kept fixed at $T=300$ K  and $\Gamma_{L,X}=\Gamma_{R,\bar{X}}=250\,\textrm{meV}, \Gamma_{R,X}=\Gamma_{L,\bar{X}}=0$, respectively. }
\end{figure}

The interpretation of the individual contributions is similar to the single-mode case. In Fig.~\ref{fig:fig7} two different experimental 
curves are shown together with the corresponding theoretical $I$-$V$
plots. Taking into account the simplicity of the model presented in
this paper, the agreement is rather good. The values used for the
charge-vibron coupling  ($\lambda_1=15(35) \, \textrm{meV},
\lambda_2=15(20)  \, \textrm{meV}$) and vibron frequencies
($\Omega_1=20 \, \textrm{meV}, \Omega_2=6 \, \textrm{meV}$)  for the
yellow (black) theoretical curves have    reasonable orders of
magnitude for low-frequency modes, see e.g.~Ref.~[\onlinecite{star05}]. We stress, however, that the  absolute values of the current  are mainly determined in our model by the size of the electronic hopping integrals; the influence of the vibrons is to modify the {\it shape} and {\it slope} of the curves. 

\ \\
To conclude, we have investigated in this paper signatures of electron-vibron interaction in the $I$-$V$ characteristics of a DNA model. Our main motivation were  recent experiments on short suspended DNA oligomers with a complex base-pair sequence.~\cite{cohen05,cohen06} The complexity of the physical system under investigation  does not allow to draw a definitive conclusion about the mechanism(s) leading to  the observed high  currents. We have shown that  vibrons coupled to the total electronic charge density  can considerably influence the current outside the zero-current gap. The ``quality'' of the molecule-electrode coupling  was also shown to  modify the orders of magnitude of the current. Another critical parameter in this model, the electronic hopping, may be  modified by non-local electron-vibron coupling related, e.g.,~to inter-base vibrations~\cite{berlin01,hennig04} or by electron-electron interactions~\cite{yi03} and as a result, the current profile  is also expected to be modify. 

Finally we would like to comment on a recent estimation of the maximum current which could  be attained in a DNA molecule, which was based on a kinetic model for a molecular  wire.~\cite{jortner05} This approach assumes  thermal hopping, i.e.~sequential tunneling  with complete destruction of the phase coherence previous to each hopping  process between nearest neighbors. Strikingly, the authors predict a maximum current of the order of {\it pico-Amperes}, in contrast to recent experimental results.~\cite{tao04,cohen05,cohen06} 
As shown in the present paper, the absolute value of the current can
be dramatically changed by  varying  the electronic hopping integrals
as well as by the way the two strands are contacted to the electrodes.
Moreover, since the electronic matrix elements used in our
investigation are on the average larger than the polaron localization
energy $\sim g^2\Omega$, we are not working in the  purely incoherent
hopping limit, where the former quantities can be treated as a small
perturbation and golden-rule-like expressions do hold. In this respect
our model differs  from the approach in Ref.~[\onlinecite{jortner05}].  Additional theoretical work is  required to bridge  kinetic and microscopic model approaches as well as to obtain reliable estimates of the electronic parameters in specific DNA wires {\it including} structural fluctuation effects.~\cite{voityuk01,siebbeles05}
 From the experimental point of view it would be highly desirable:
 (i) to perform a systematic study on the effect of base pair sequence
 and length dependence on the current and the conductance within the
 set-up of Ref.~[\onlinecite{cohen05,cohen06}], since  the length-scaling of the linear conductance is an important benchmark for disclosing the most effective transport channels in molecular wires; (ii) to explore different contact geometries. 
 
\section{Acknowledgments}
We  acknowledge fruitful discussions with Igor Brodsky, Joshua Jortner, Ron Naaman, Claude Nogues, and Dmitri Ryndyk. This work has been supported by the Volkswagen Foundation grant Nr.~I/78-340, by the EU under contract IST-2001-38951 and by the Israeli Academy of Sciences and Humanities.

\begin{appendix} 
\section{Derivation of Eq.~(\ref{eq:eqcur})}
In order to derive Eq.~\ref{eq:eqcur}, all  we need to show is that 
terms proportional to $\textrm{Tr}\{\Gamma_{\rml}\Gamma_{\rml} \cdots \}$ and
$\textrm{Tr}\{\Gamma_{\rmr}\Gamma_{\rmr}\cdots \}$ identically vanish. Let's assume e.~g.~, that 
$\Gamma_{\rmr}=0$. Thus, the left-going current, which is proportional to
$\Gamma_{\rml}\Gamma_{\rml}$  {\em must} be zero. From Eq.~\ref{eq:eq2}, we first
obtain:
\begin{eqnarray}
J_{\rml}&=&\frac{2\ii e}{h} \sum^{\infty}_{n=-\infty}\phi_n(\tau)\, \int dE \; \textrm{Tr}[ \Gamma_{\rml} \{ f_{\rml}(E)
G^{>}(E-n\Omega) \nonumber \\ &+& (1-f_{\rml}(E)) G^{<}(E+n\Omega)   \}].
\end{eqnarray}
Use now the kinetic equation for the Green function
$G^{<(>)}(E)=G^{r}(E)\Sigma^{<(>)}_{\rml}(E)G^{a}(E)$, where $\Gamma_{\rmr}=0$
has been already set, and  insert it in the above equation with the short-hand
notations $C_{\pm}=\textrm{Tr}\{\Gamma_{\rml}G^{r}(E\pm n\Omega) \Gamma_{\rml}G^{a}(E\pm
n\Omega) \}$.  We get: 
\begin{eqnarray}
J_{\rml}&=&\frac{2 e}{h} \sum^{\infty}_{n=-\infty}\phi_n(\tau)\, \int dE \{ 
f_{\rml}(E)\,(1-f_{\rml}(E-n\Omega)) C_{-} \nonumber  \\ &-& (1-f_{\rml}(E)) \,
f_{\rml}(E+n\Omega) C_{+} \}.
\end{eqnarray}
If we now change $n\to -n$ in the second term and use the symmetry
$\phi_{-n}=\phi_{n}e^{-\beta n\Omega}$ together with the identity
$(1-f_{\rml}(E))=e^{\beta (E-\mu_{\rml})}f_{\rml}(E)$, we find: 
 \begin{eqnarray}
J_{\rml}&=&\frac{2 e}{h} \sum^{\infty}_{n=-\infty}\phi_n(\tau)\, \int dE \, 
f_{\rml}(E)\,(1-f_{\rml}(E-n\Omega)) C_{-}  \nonumber \\ &\times &\{ 1-e^{\beta (E-\mu_{\rml})}
e^{-\beta (E-\mu_{\rml}-n\Omega)} e^{-\beta n\Omega} \}=0.
\end{eqnarray}
This means that only mixed terms containing  $\textrm{Tr}\{\Gamma_{\rml}\cdots
\Gamma_{\rmr}\}$ contribute to the current. It is then straightforward to show
along the same lines that Eq.~\ref{eq:eqcur} comes out.

\section{Asymptotic behavior of the current as a function of  $\lambda$ and $\Omega$}

The asymptotic behavior of the current as a function of the electron-vibron coupling can be immediately understood by looking at the prefactors $\phi_{n}(\tau)$, since only at this place $\lambda$ does appear. Using the asymptotic 
behavior of the Bessel functions, $I_{n}(z\ll 1)\sim z^{n}/2^{n}n!$ and $I_{n}(z\gg 1)\sim e^{z}/\sqrt{2\pi z}$, one sees that:
 \begin{eqnarray}
 \phi_{n}(g\ll 1)&\sim &e^{-g^2(2N+1)} \, e^{\beta\Omega n/2} \, \frac{g^{2n}}{n!} [N(N+1)]^{n/2} \nonumber \\ 
 &\sim& e^{-g^2(2N+1)}  g^{2n}, \nonumber \\
 \phi_{n}(g\gg 1)&\sim & e^{-g^2(2N+1)} \, e^{\beta\Omega n/2} \, \frac{e^{2g^2Ne^{\beta\Omega/2}} } {\sqrt{4\pi g^2 Ne^{\beta\Omega/2} }} \nonumber \\
 &=& \frac{e^{\beta\Omega n/2} }{\sqrt{4\pi g^2 Ne^{\beta\Omega/2} }} \, 
 e^{-g^2 \, \overbrace{N [ e^{\beta\Omega}-2e^{\beta\Omega/2}+1] }^{A >0 !}} ,\nonumber \\
 \phi_{n}(g\gg 1)&\sim& \frac{e^{-g^2 A}}{g} \to  0. \nonumber
 \end{eqnarray}
From here it follows that the inelastic current will grow as some power of $g^2$ and then decay to zero, while the elastic part ($n=0$) starts from its bare value (at $\lambda=0$) and then rapidly decays for larger values of the electron-vibron interaction.

In order to analyze the behavior of the current for small and large frequencies, it is appropriate to write Eq.~\ref{eq:eqcur} in a slightly different form. Doing a change of variables in the $+$ and $-$ components, we arrive at: 

  \begin{eqnarray}
  j_{\textrm{T}}&=&
\phi_{0}(\tau)\, 2\frac{e}{h}\,\int dE \, t(E) \, (f_{\rml}(E)-f_{\rmr}(E))+ \nonumber \\
 &+& \frac{e}{h}\,\sum_{n\neq 0} \phi_{n}(\tau) \, \int dE \, t(E) \, \Lambda_{n}(E,\Omega) \nonumber \\
&=& j_{\textrm{el}}+ j_{\textrm{inel}}\nonumber\\
  \Lambda_{n}(E,\Omega) &=& f_{\rml}(E+n\Omega) (1-f_{\rmr}(E)) - \nonumber \\ &-& f_{\rmr}(E+n\Omega) (1-f_{\rml}(E)) \nonumber \\
&+&
 f_{\rml}(E) (1-f_{\rmr}(E-n\Omega)) - \nonumber \\ &-& f_{\rmr}(E) (1-f_{\rml}(E-n\Omega)). \nonumber
  \end{eqnarray}
The sum can be now split into terms with $n >0$ and terms with $n<0$. Using the symmetry  $\phi_{-n}(\tau)=\phi_{n}(\tau)e^{-\beta\Omega n}$, the above expressions can be cast into the following form:
\begin{eqnarray}
&& j_{\textrm{T}}=
\phi_{0}(\tau)\, 2\frac{e}{h}\,\int dE \, t(E) \, (f_{\rml}(E)-f_{\rmr}(E))+\nonumber\\
&+&  \frac{e}{h}\,\sum_{n\ge 1} \phi_{n}(\tau)  \int dE \, t(E) \left\lbrace  \Lambda_{n}(E,\Omega) + \Lambda_{-n}(E,\Omega)  e^{-\beta\Omega n} \right\rbrace \nonumber 
\end{eqnarray}
Let's   consider the case  of large vibron frequency.  We can use the fact that $f(E\pm n\Omega)$ goes to 0 ($+$) or 1 ($-$) when $\Omega\to\infty$. Note that in this case $\Lambda_{n}(E,\Omega)$ vanishes while $\Lambda_{-n}(E,\Omega)$ goes over into $2(f_{\rml}(E)-f_{\rmr}(E))$. Using this result together with the asymptotic behavior  of the Bessel functions for small arguments leads to:
\begin{eqnarray}
j_{\textrm{inel}}&\approx&2\frac{e}{h}
 \sum_{n\ge 1} \int dE \, t(E) \, (f_{\rml}(E)-f_{\rmr}(E))\, \times \nonumber \\
&\times& e^{-g^2}\frac{g^{2n}}{n!} \, e^{-\beta\Omega n/2}\,e^{\beta\Omega n/2} e^{-\beta\Omega n}
\stackrel{\Omega\to\infty}{\longrightarrow} 0\nonumber 
\end{eqnarray}
Hence, the inelastic current vanishes at very large frequencies. The elastic current, however, saturates at the value $(2e/h)\int dE \, t(E) (f_{\textrm{L}}(E)-f_{\textrm{R}}(E))$, since $\phi_{n=0}(\tau)\sim e^{-g^2}I_{0}(2g^2e^{-\beta\Omega})\to e^{-g^2}\to 1$ when $\Omega\to\infty$.

 In the case $\Omega\to 0$, the inelastic part of the current will adopt the following form (with $x=\hbar\Omega/k_{\textrm{B}}T)$: 
 \begin{eqnarray}
 j_{\textrm{inel}}&\approx& 4\frac{e}{h} \sum_{n\ge 1} \int dE \, t(E) \, (f_{\rml}(E)-f_{\rmr}(E)) \, \times \nonumber \\ &\times & e^{-2g^2/x} 
 \frac{e^{2g^2/x}}{\sqrt{4\pi g^2}}\, \sqrt{x} \,e^{\beta\Omega n} \sim  \sqrt{x} \stackrel{x\to 0}{\longrightarrow} 0 , \nonumber
 \end{eqnarray}
 where the asymptotic expansion of the Bessel functions for large argument has been used. 
A similar scaling would follow for the elastic part of the current, so that for $\Omega\to 0$ the total current is suppressed. This is clearly an artifact of the limiting procedure, since the Lang-Firsov transformation is obviously ill-defined at zero frequency.

\end{appendix}
%


\begin{thebibliography}{46}
\expandafter\ifx\csname natexlab\endcsname\relax\def\natexlab#1{#1}\fi
\expandafter\ifx\csname bibnamefont\endcsname\relax
  \def\bibnamefont#1{#1}\fi
\expandafter\ifx\csname bibfnamefont\endcsname\relax
  \def\bibfnamefont#1{#1}\fi
\expandafter\ifx\csname citenamefont\endcsname\relax
  \def\citenamefont#1{#1}\fi
\expandafter\ifx\csname url\endcsname\relax
  \def\url#1{\texttt{#1}}\fi
\expandafter\ifx\csname urlprefix\endcsname\relax\def\urlprefix{URL }\fi
\providecommand{\bibinfo}[2]{#2}
\providecommand{\eprint}[2][]{\url{#2}}

\bibitem[{\citenamefont{Cuniberti et~al.}(2005)\citenamefont{Cuniberti, Fagas,
  and (Eds.)}}]{ME05}
\bibinfo{author}{\bibfnamefont{G.}~\bibnamefont{Cuniberti}},
  \bibinfo{author}{\bibfnamefont{G.}~\bibnamefont{Fagas}}, \bibnamefont{and}
  \bibinfo{author}{\bibfnamefont{K.~R.} \bibnamefont{(Eds.)}},
  \emph{\bibinfo{title}{Introducing Molecular Electronics: A brief overview}} (\bibinfo{publisher}{Springer}), 
  \bibinfo{journal}{Lecture Notes in Physics} \textbf{\bibinfo{volume}{680}}
  (\bibinfo{year}{2005}).



\bibitem[{\citenamefont{Keren et~al.}(2003)\citenamefont{Keren, Berman,
  Buchstab, Sivan, and Braun}}]{keren03}
\bibinfo{author}{\bibfnamefont{K.}~\bibnamefont{Keren}},
  \bibinfo{author}{\bibfnamefont{R.~S.} \bibnamefont{Berman}},
  \bibinfo{author}{\bibfnamefont{E.}~\bibnamefont{Buchstab}},
  \bibinfo{author}{\bibfnamefont{U.}~\bibnamefont{Sivan}}, \bibnamefont{and}
  \bibinfo{author}{\bibfnamefont{E.}~\bibnamefont{Braun}},
  \bibinfo{journal}{Science} \textbf{\bibinfo{volume}{302}},
  \bibinfo{pages}{1380} (\bibinfo{year}{2003}).

\bibitem[{\citenamefont{Mertig et~al.}(1999)\citenamefont{Mertig, Kirsch,
  Pompe, and Engelhardt}}]{pompe99}
\bibinfo{author}{\bibfnamefont{M.}~\bibnamefont{Mertig}},
  \bibinfo{author}{\bibfnamefont{R.}~\bibnamefont{Kirsch}},
  \bibinfo{author}{\bibfnamefont{W.}~\bibnamefont{Pompe}}, \bibnamefont{and}
  \bibinfo{author}{\bibfnamefont{H.}~\bibnamefont{Engelhardt}},
  \bibinfo{journal}{Eur. Phys. J. D} \textbf{\bibinfo{volume}{9}},
  \bibinfo{pages}{45} (\bibinfo{year}{1999}).

\bibitem[{\citenamefont{Treadway et~al.}(2002)\citenamefont{Treadway, Hill, and
  Barton}}]{treadway02}
\bibinfo{author}{\bibfnamefont{C.~R.} \bibnamefont{Treadway}},
  \bibinfo{author}{\bibfnamefont{M.~G.} \bibnamefont{Hill}}, \bibnamefont{and}
  \bibinfo{author}{\bibfnamefont{J.~K.} \bibnamefont{Barton}},
  \bibinfo{journal}{Chem.\ Phys.} \textbf{\bibinfo{volume}{281}},
  \bibinfo{pages}{409} (\bibinfo{year}{2002}).

\bibitem[{\citenamefont{Murphy et~al.}(1993)\citenamefont{Murphy, Arkin,
  Jenkins, Ghatlia, Bossmann, Turro, and Barton}}]{murphy93}
\bibinfo{author}{\bibfnamefont{C.~J.} \bibnamefont{Murphy}},
  \bibinfo{author}{\bibfnamefont{M.~R.} \bibnamefont{Arkin}},
  \bibinfo{author}{\bibfnamefont{Y.}~\bibnamefont{Jenkins}},
  \bibinfo{author}{\bibfnamefont{N.~D.} \bibnamefont{Ghatlia}},
  \bibinfo{author}{\bibfnamefont{S.~H.} \bibnamefont{Bossmann}},
  \bibinfo{author}{\bibfnamefont{N.~J.} \bibnamefont{Turro}}, \bibnamefont{and}
  \bibinfo{author}{\bibfnamefont{J.~K.} \bibnamefont{Barton}},
  \bibinfo{journal}{Science} \textbf{\bibinfo{volume}{262}},
  \bibinfo{pages}{1025} (\bibinfo{year}{1993}).

\bibitem[{\citenamefont{Braun et~al.}(1998)\citenamefont{Braun, Eichen, Sivan,
  and Ben-Yoseph}}]{braun98}
\bibinfo{author}{\bibfnamefont{E.}~\bibnamefont{Braun}},
  \bibinfo{author}{\bibfnamefont{Y.}~\bibnamefont{Eichen}},
  \bibinfo{author}{\bibfnamefont{U.}~\bibnamefont{Sivan}}, \bibnamefont{and}
  \bibinfo{author}{\bibfnamefont{G.}~\bibnamefont{Ben-Yoseph}},
  \bibinfo{journal}{Nature} \textbf{\bibinfo{volume}{391}},
  \bibinfo{pages}{775} (\bibinfo{year}{1998}).

\bibitem[{\citenamefont{Storm et~al.}(2001)\citenamefont{Storm, Noort, Vries,
  and Dekker}}]{storm01}
\bibinfo{author}{\bibfnamefont{A.~J.} \bibnamefont{Storm}},
  \bibinfo{author}{\bibfnamefont{J.~V.} \bibnamefont{Noort}},
  \bibinfo{author}{\bibfnamefont{S.~D.} \bibnamefont{Vries}}, \bibnamefont{and}
  \bibinfo{author}{\bibfnamefont{C.}~\bibnamefont{Dekker}},
  \bibinfo{journal}{Appl.\ Phys.\ Lett.} \textbf{\bibinfo{volume}{79}},
  \bibinfo{pages}{3881} (\bibinfo{year}{2001}).

\bibitem[{\citenamefont{Porath et~al.}(2000)\citenamefont{Porath, Bezryadin,
  Vries, and Dekker}}]{porath00}
\bibinfo{author}{\bibfnamefont{D.}~\bibnamefont{Porath}},
  \bibinfo{author}{\bibfnamefont{A.}~\bibnamefont{Bezryadin}},
  \bibinfo{author}{\bibfnamefont{S.~D.} \bibnamefont{Vries}}, \bibnamefont{and}
  \bibinfo{author}{\bibfnamefont{C.}~\bibnamefont{Dekker}},
  \bibinfo{journal}{Nature} \textbf{\bibinfo{volume}{403}},
  \bibinfo{pages}{635} (\bibinfo{year}{2000}).

\bibitem[{\citenamefont{Yoo et~al.}(2001)\citenamefont{Yoo, Ha, Lee, Park, Kim,
  Kim, Lee, Kawai, and Choi}}]{yoo01}
\bibinfo{author}{\bibfnamefont{K.-H.} \bibnamefont{Yoo}},
  \bibinfo{author}{\bibfnamefont{D.~H.} \bibnamefont{Ha}},
  \bibinfo{author}{\bibfnamefont{J.-O.} \bibnamefont{Lee}},
  \bibinfo{author}{\bibfnamefont{J.~W.} \bibnamefont{Park}},
  \bibinfo{author}{\bibfnamefont{J.}~\bibnamefont{Kim}},
  \bibinfo{author}{\bibfnamefont{J.~J.} \bibnamefont{Kim}},
  \bibinfo{author}{\bibfnamefont{H.-Y.} \bibnamefont{Lee}},
  \bibinfo{author}{\bibfnamefont{T.}~\bibnamefont{Kawai}}, \bibnamefont{and}
  \bibinfo{author}{\bibfnamefont{H.~Y.} \bibnamefont{Choi}},
  \bibinfo{journal}{Phys.\ Rev.\ Lett.} \textbf{\bibinfo{volume}{87}},
   \bibinfo{pages}{198102}  (\bibinfo{year}{2001}).

\bibitem[{\citenamefont{Xu et~al.}(2004)\citenamefont{Xu, Zhang, Li, and
  Tao}}]{tao04}
\bibinfo{author}{\bibfnamefont{B.}~\bibnamefont{Xu}},
  \bibinfo{author}{\bibfnamefont{P.}~\bibnamefont{Zhang}},
  \bibinfo{author}{\bibfnamefont{X.}~\bibnamefont{Li}}, \bibnamefont{and}
  \bibinfo{author}{\bibfnamefont{N.}~\bibnamefont{Tao}}, \bibinfo{journal}{Nano
  letters} \textbf{\bibinfo{volume}{4}}, \bibinfo{pages}{1105}
  (\bibinfo{year}{2004}).

\bibitem[{\citenamefont{Cohen et~al.}(2005)\citenamefont{Cohen, Nogues, Naaman,
  and Porath}}]{cohen05}
\bibinfo{author}{\bibfnamefont{H.}~\bibnamefont{Cohen}},
  \bibinfo{author}{\bibfnamefont{C.}~\bibnamefont{Nogues}},
  \bibinfo{author}{\bibfnamefont{R.}~\bibnamefont{Naaman}}, \bibnamefont{and}
  \bibinfo{author}{\bibfnamefont{D.}~\bibnamefont{Porath}},
  \bibinfo{journal}{Proc.\ Natl.\ Acad.\ Sci.\ USA}
  \textbf{\bibinfo{volume}{102}}, \bibinfo{pages}{11589}
  (\bibinfo{year}{2005}).


\bibitem[{\citenamefont{Cohen et~al.}(2006)\citenamefont{Cohen, Nogues, Ullien, Daube, Naaman,
  and Porath}}]{cohen06}
\bibinfo{author}{\bibfnamefont{H.}~\bibnamefont{Cohen}},
  \bibinfo{author}{\bibfnamefont{C.}~\bibnamefont{Nogues}},
 \bibinfo{author}{\bibfnamefont{D.}~\bibnamefont{Ullien}},
\bibinfo{author}{\bibfnamefont{S.}~\bibnamefont{Daube}},
  \bibinfo{author}{\bibfnamefont{R.}~\bibnamefont{Naaman}}, \bibnamefont{and}
  \bibinfo{author}{\bibfnamefont{D.}~\bibnamefont{Porath}},
  \bibinfo{journal}{Faraday Discussions}
  \textbf{\bibinfo{volume}{131}}, \bibinfo{pages}{367}
  (\bibinfo{year}{2006}).



\bibitem[{\citenamefont{Cohen et~al.}(2004)\citenamefont{Nogues,Cohen, Daube, and Naaman
  }}]{cohen04}
  \bibinfo{author}{\bibfnamefont{C.}~\bibnamefont{Nogues}},
\bibinfo{author}{\bibfnamefont{H.}~\bibnamefont{Cohen}},
\bibinfo{author}{\bibfnamefont{S.}~\bibnamefont{Daube}}, \bibnamefont{and}
  \bibinfo{author}{\bibfnamefont{R.}~\bibnamefont{Naaman}}, 
  \bibinfo{journal}{Phys.\ Chem.\ Chem. \ Phys.}
  \textbf{\bibinfo{volume}{6}}, \bibinfo{pages}{4459}
  (\bibinfo{year}{2004}).



\bibitem[{\citenamefont{Artacho et~al.}(2003)\citenamefont{Artacho, Machado,
  Sanchez-Portal, Ordejon, and Soler}}]{artacho03}
\bibinfo{author}{\bibfnamefont{E.}~\bibnamefont{Artacho}},
  \bibinfo{author}{\bibfnamefont{M.}~\bibnamefont{Machado}},
  \bibinfo{author}{\bibfnamefont{D.}~\bibnamefont{Sanchez-Portal}},
  \bibinfo{author}{\bibfnamefont{P.}~\bibnamefont{Ordejon}}, \bibnamefont{and}
  \bibinfo{author}{\bibfnamefont{J.~M.} \bibnamefont{Soler}},
  \bibinfo{journal}{Mol. Phys.} \textbf{\bibinfo{volume}{101}},
  \bibinfo{pages}{1587} (\bibinfo{year}{2003}).

\bibitem[{\citenamefont{Calzolari et~al.}(2002)\citenamefont{Calzolari, Felice,
  Molinari, and Garbesi}}]{difelice02}
\bibinfo{author}{\bibfnamefont{A.}~\bibnamefont{Calzolari}},
  \bibinfo{author}{\bibfnamefont{R.~D.} \bibnamefont{Felice}},
  \bibinfo{author}{\bibfnamefont{E.}~\bibnamefont{Molinari}}, \bibnamefont{and}
  \bibinfo{author}{\bibfnamefont{A.}~\bibnamefont{Garbesi}},
  \bibinfo{journal}{Appl.\ Phys.\ Lett.} \textbf{\bibinfo{volume}{80}},
  \bibinfo{pages}{3331} (\bibinfo{year}{2002}).

\bibitem[{\citenamefont{Felice et~al.}(2004)\citenamefont{Felice, Calzolari,
  and Zhang}}]{felice04}
\bibinfo{author}{\bibfnamefont{R.~D.} \bibnamefont{Felice}},
  \bibinfo{author}{\bibfnamefont{A.}~\bibnamefont{Calzolari}},
  \bibnamefont{and} \bibinfo{author}{\bibfnamefont{H.}~\bibnamefont{Zhang}},
  \bibinfo{journal}{Nanotechnology} \textbf{\bibinfo{volume}{15}},
  \bibinfo{pages}{1256} (\bibinfo{year}{2004}).

\bibitem[{\citenamefont{Gervasio et~al.}(2002)\citenamefont{Gervasio, Carloni,
  and Parrinello}}]{gervasio02}
\bibinfo{author}{\bibfnamefont{F.~L.} \bibnamefont{Gervasio}},
  \bibinfo{author}{\bibfnamefont{P.}~\bibnamefont{Carloni}}, \bibnamefont{and}
  \bibinfo{author}{\bibfnamefont{M.}~\bibnamefont{Parrinello}},
  \bibinfo{journal}{Phys.\ Rev.\ Lett.} \textbf{\bibinfo{volume}{89}},
  \bibinfo{pages}{108102} (\bibinfo{year}{2002}).

\bibitem[{\citenamefont{Barnett et~al.}(2001)\citenamefont{Barnett, Cleveland,
  Joy, Landman, and Schuster}}]{barnett01}
\bibinfo{author}{\bibfnamefont{R.~N.} \bibnamefont{Barnett}},
  \bibinfo{author}{\bibfnamefont{C.~L.} \bibnamefont{Cleveland}},
  \bibinfo{author}{\bibfnamefont{A.}~\bibnamefont{Joy}},
  \bibinfo{author}{\bibfnamefont{U.}~\bibnamefont{Landman}}, \bibnamefont{and}
  \bibinfo{author}{\bibfnamefont{G.~B.} \bibnamefont{Schuster}},
  \bibinfo{journal}{Science} \textbf{\bibinfo{volume}{294}},
  \bibinfo{pages}{567} (\bibinfo{year}{2001}).

\bibitem[{\citenamefont{H{\"u}bsch et~al.}(2005)\citenamefont{H{\"u}bsch,
  Endres, Cox, and Singh}}]{endres05}
\bibinfo{author}{\bibfnamefont{A.}~\bibnamefont{H{\"u}bsch}},
  \bibinfo{author}{\bibfnamefont{R.~G.} \bibnamefont{Endres}},
  \bibinfo{author}{\bibfnamefont{D.~L.} \bibnamefont{Cox}}, \bibnamefont{and}
  \bibinfo{author}{\bibfnamefont{R.~R.~P.} \bibnamefont{Singh}},
  \bibinfo{journal}{Phys.\ Rev.\ Lett.} \textbf{\bibinfo{volume}{94}},
  \bibinfo{pages}{178102} (\bibinfo{year}{2005}).

\bibitem[{\citenamefont{Starikov}(2003)}]{star04}
\bibinfo{author}{\bibfnamefont{E.~B.} \bibnamefont{Starikov}},
  \bibinfo{journal}{Phil. Mag. Lett.} \textbf{\bibinfo{volume}{83}},
  \bibinfo{pages}{699} (\bibinfo{year}{2003}).

\bibitem[{\citenamefont{Starikov}(2005)}]{star05}
\bibinfo{author}{\bibfnamefont{E.~B.} \bibnamefont{Starikov}},
  \bibinfo{journal}{Phil. Mag.} \textbf{\bibinfo{volume}{85}},
  \bibinfo{pages}{3435} (\bibinfo{year}{2005}).

\bibitem[{\citenamefont{Adessi et~al.}(2003)\citenamefont{Adessi, Walch, and
  Anantram}}]{adessi03}
\bibinfo{author}{\bibfnamefont{C.}~\bibnamefont{Adessi}},
  \bibinfo{author}{\bibfnamefont{S.}~\bibnamefont{Walch}}, \bibnamefont{and}
  \bibinfo{author}{\bibfnamefont{M.~P.} \bibnamefont{Anantram}},
  \bibinfo{journal}{Phys.\ Rev.\ B} \textbf{\bibinfo{volume}{67}},
  \bibinfo{pages}{081405(R)} (\bibinfo{year}{2003}).

\bibitem[{\citenamefont{Mehrez and Anantram}(2005)}]{mehrez05}
\bibinfo{author}{\bibfnamefont{H.}~\bibnamefont{Mehrez}} \bibnamefont{and}
  \bibinfo{author}{\bibfnamefont{M.~P.} \bibnamefont{Anantram}},
  \bibinfo{journal}{Phys.\ Rev.\ B} \textbf{\bibinfo{volume}{71}},
  \bibinfo{pages}{115405} (\bibinfo{year}{2005}).

\bibitem[{\citenamefont{Cuniberti et~al.}(2002)\citenamefont{Cuniberti, Craco,
  Porath, and Dekker}}]{gio02}
\bibinfo{author}{\bibfnamefont{G.}~\bibnamefont{Cuniberti}},
  \bibinfo{author}{\bibfnamefont{L.}~\bibnamefont{Craco}},
  \bibinfo{author}{\bibfnamefont{D.}~\bibnamefont{Porath}}, \bibnamefont{and}
  \bibinfo{author}{\bibfnamefont{C.}~\bibnamefont{Dekker}},
  \bibinfo{journal}{Phys.\ Rev.\ B} \textbf{\bibinfo{volume}{65}},
  \bibinfo{pages}{241314(R)} (\bibinfo{year}{2002}).

\bibitem[{\citenamefont{Jortner et~al.}(1998)\citenamefont{Jortner, Bixon,
  Langenbacher, and Michel-Beyerle}}]{jortner98}
\bibinfo{author}{\bibfnamefont{J.}~\bibnamefont{Jortner}},
  \bibinfo{author}{\bibfnamefont{M.}~\bibnamefont{Bixon}},
  \bibinfo{author}{\bibfnamefont{T.}~\bibnamefont{Langenbacher}},
  \bibnamefont{and}
  \bibinfo{author}{\bibfnamefont{M.}~\bibnamefont{Michel-Beyerle}},
  \bibinfo{journal}{Proc. Natl. Acad. Sci.} \textbf{\bibinfo{volume}{95}},
  \bibinfo{pages}{12759} (\bibinfo{year}{1998}).

\bibitem[{\citenamefont{Jortner and Bixon}(2002)}]{jortner02}
\bibinfo{author}{\bibfnamefont{J.}~\bibnamefont{Jortner}} \bibnamefont{and}
  \bibinfo{author}{\bibfnamefont{M.}~\bibnamefont{Bixon}},
  \bibinfo{journal}{Chemical Physics} \textbf{\bibinfo{volume}{281}},
  \bibinfo{pages}{393} (\bibinfo{year}{2002}).

\bibitem[{\citenamefont{Roche et~al.}(2003)\citenamefont{Roche, Bicout, Macia,
  and Kats}}]{roche03a}
\bibinfo{author}{\bibfnamefont{S.}~\bibnamefont{Roche}},
  \bibinfo{author}{\bibfnamefont{D.}~\bibnamefont{Bicout}},
  \bibinfo{author}{\bibfnamefont{E.}~\bibnamefont{Macia}}, \bibnamefont{and}
  \bibinfo{author}{\bibfnamefont{E.}~\bibnamefont{Kats}},
  \bibinfo{journal}{Phys.\ Rev.\ Lett.} \textbf{\bibinfo{volume}{91}},
  \bibinfo{pages}{228101} (\bibinfo{year}{2003}).

\bibitem[{\citenamefont{Roche}(2003)}]{roche03}
\bibinfo{author}{\bibfnamefont{S.}~\bibnamefont{Roche}},
  \bibinfo{journal}{Phys.\ Rev.\ Lett.} \textbf{\bibinfo{volume}{91}},
  \bibinfo{pages}{108101} (\bibinfo{year}{2003}).


\bibitem[{\citenamefont{Unge and Stafstrom}(2003)}]{unge03}
\bibinfo{author}{\bibfnamefont{M.}~\bibnamefont{Unge}} \bibnamefont{and}
  \bibinfo{author}{\bibfnamefont{S.}~\bibnamefont{Stafstrom}},
  \bibinfo{journal}{Nano letters} \textbf{\bibinfo{volume}{3}},
  \bibinfo{pages}{1417} (\bibinfo{year}{2003}).

\bibitem[{\citenamefont{Hennig et~al.}(2004)\citenamefont{Hennig, Starikov,
  Archilla, and Palmero}}]{hennig04a}
\bibinfo{author}{\bibfnamefont{D.}~\bibnamefont{Hennig}},
  \bibinfo{author}{\bibfnamefont{E.~B.} \bibnamefont{Starikov}},
  \bibinfo{author}{\bibfnamefont{J.~F.~R.} \bibnamefont{Archilla}},
  \bibnamefont{and} \bibinfo{author}{\bibfnamefont{F.}~\bibnamefont{Palmero}},
  \bibinfo{journal}{Journal of Biological Physics}
  \textbf{\bibinfo{volume}{30}}, \bibinfo{pages}{227} (\bibinfo{year}{2004}).

\bibitem[{\citenamefont{Jortner and Bixon}(2005)}]{jortner05}
\bibinfo{author}{\bibfnamefont{J.}~\bibnamefont{Jortner}} \bibnamefont{and}
  \bibinfo{author}{\bibfnamefont{M.}~\bibnamefont{Bixon}},
  \bibinfo{journal}{Chem. Phys.} \textbf{\bibinfo{volume}{319}},
  \bibinfo{pages}{273} (\bibinfo{year}{2005}).

\bibitem[{\citenamefont{Apalkov and Chakraborty}(2005{\natexlab{a}})}]{ac05}
\bibinfo{author}{\bibfnamefont{V.~M.} \bibnamefont{Apalkov}} \bibnamefont{and}
  \bibinfo{author}{\bibfnamefont{T.}~\bibnamefont{Chakraborty}},
  \bibinfo{journal}{Phys.\ Rev.\ B} \textbf{\bibinfo{volume}{71}},
  \bibinfo{pages}{033102} (\bibinfo{year}{2005}{\natexlab{a}}).

\bibitem[{\citenamefont{Apalkov and Chakraborty}(2005{\natexlab{b}})}]{ac05a}
\bibinfo{author}{\bibfnamefont{V.} \bibnamefont{Apalkov}} \bibnamefont{and}
  \bibinfo{author}{\bibfnamefont{T.}~\bibnamefont{Chakraborty}},
  \bibinfo{journal}{Phys.\ Rev.\ B} \textbf{\bibinfo{volume}{72}},
  \bibinfo{pages}{161102(R)} (\bibinfo{year}{2005}{\natexlab{b}}).

\bibitem[{\citenamefont{Gutierrez
  et~al.}(2005{\natexlab{a}})\citenamefont{Gutierrez, Mandal, and
  Cuniberti}}]{gmc05a}
\bibinfo{author}{\bibfnamefont{R.}~\bibnamefont{Gutierrez}},
  \bibinfo{author}{\bibfnamefont{S.}~\bibnamefont{Mandal}}, \bibnamefont{and}
  \bibinfo{author}{\bibfnamefont{G.}~\bibnamefont{Cuniberti}},
  \bibinfo{journal}{Nano letters} \textbf{\bibinfo{volume}{5}},
  \bibinfo{pages}{1093} (\bibinfo{year}{2005}{\natexlab{a}}).

\bibitem[{\citenamefont{Gutierrez
  et~al.}(2005{\natexlab{b}})\citenamefont{Gutierrez, Mandal, and
  Cuniberti}}]{gmc05b}
\bibinfo{author}{\bibfnamefont{R.}~\bibnamefont{Gutierrez}},
  \bibinfo{author}{\bibfnamefont{S.}~\bibnamefont{Mandal}}, \bibnamefont{and}
  \bibinfo{author}{\bibfnamefont{G.}~\bibnamefont{Cuniberti}},
  \bibinfo{journal}{Phys.\ Rev.\ B} \textbf{\bibinfo{volume}{71}},
  \bibinfo{pages}{235116} (\bibinfo{year}{2005}{\natexlab{b}}).

 \bibitem[{\citenamefont{Kohler et~al.}(2005)\citenamefont{Kohler, Lehmann, and
   H{\"a}nggi}}]{klh05}
 \bibinfo{author}{\bibfnamefont{S.}~\bibnamefont{Kohler}},
   \bibinfo{author}{\bibfnamefont{J.}~\bibnamefont{Lehmann}}, \bibnamefont{and}
   \bibinfo{author}{\bibfnamefont{P.}~\bibnamefont{H{\"a}nggi}},
   \bibinfo{journal}{Phys. Rep.} \textbf{\bibinfo{volume}{406}},
   \bibinfo{pages}{379} (\bibinfo{year}{2005}).

\bibitem[{\citenamefont{Yamada et~al.}(2004)\citenamefont{Yamada, Starikov,
  Hennig, and Archilla}}]{yamada04a}
\bibinfo{author}{\bibfnamefont{H.}~\bibnamefont{Yamada}},
  \bibinfo{author}{\bibfnamefont{E.~B.} \bibnamefont{Starikov}},
  \bibinfo{author}{\bibfnamefont{D.}~\bibnamefont{Hennig}}, \bibnamefont{and}
  \bibinfo{author}{\bibfnamefont{J.~F.~R.} \bibnamefont{Archilla}},
  \bibinfo{journal}{{Eur. Phys. J. E}}  \textbf{\bibinfo{volume}{17}},   \bibinfo{pages}{149}  (\bibinfo{year}{2005}).

\bibitem[{\citenamefont{D'Orsogna and Bruinsma}(2003)}]{bruinsma03}
\bibinfo{author}{\bibfnamefont{M.~R.} \bibnamefont{D'Orsogna}}
  \bibnamefont{and} \bibinfo{author}{\bibfnamefont{R.}~\bibnamefont{Bruinsma}},
  \bibinfo{journal}{Phys.\ Rev.\ Lett.} \textbf{\bibinfo{volume}{90}},
  \bibinfo{pages}{078301} (\bibinfo{year}{2003}).

\bibitem[{\citenamefont{Koch and von Oppen}(2005)}]{kvo05}
\bibinfo{author}{\bibfnamefont{J.}~\bibnamefont{Koch}} \bibnamefont{and}
  \bibinfo{author}{\bibfnamefont{F.}~\bibnamefont{von Oppen}},
  \bibinfo{journal}{Phys.\ Rev.\ Lett.} \textbf{\bibinfo{volume}{94}},
  \bibinfo{pages}{206804} (\bibinfo{year}{2005}).


\bibitem[{ \citenamefont{de Pablo et~al.}(2000)\citenamefont{dePablo, Colchero, Luna, Gomez-Herrero, and Baro}}]{pablo00}
\bibinfo{author}{\bibfnamefont{P.~J.}~\bibnamefont{dePablo}},
\bibinfo{author}{\bibfnamefont{J.}~\bibnamefont{Colchero}}, 
\bibinfo{author}{\bibfnamefont{M.}~\bibnamefont{Luna}},
\bibinfo{author}{\bibfnamefont{J.}~\bibnamefont{Gomez-Herrero}},
\bibnamefont{and}
\bibinfo{author}{\bibfnamefont{A.~M.}~\bibnamefont{Baro}},
  \bibinfo{journal}{Phys.\ Rev.\ B} \textbf{\bibinfo{volume}{61}},
  \bibinfo{pages}{14179} (\bibinfo{year}{2000}).

\bibitem[{\citenamefont{Yi}(2003)}]{yi03}
\bibinfo{author}{\bibfnamefont{J.}~\bibnamefont{Yi}}, \bibinfo{journal}{Phys.\
  Rev.\ B} \textbf{\bibinfo{volume}{68}}, \bibinfo{pages}{193103}
  (\bibinfo{year}{2003}).

\bibitem[{\citenamefont{Yamada}(2004)}]{yamada04}
\bibinfo{author}{\bibfnamefont{H.}~\bibnamefont{Yamada}},
  \bibinfo{journal}{Phys. Lett. A} \textbf{\bibinfo{volume}{332}},
  \bibinfo{pages}{65} (\bibinfo{year}{2004}).

\bibitem[{\citenamefont{Klotsa et~al.}(2005)\citenamefont{Klotsa, Roemer, and
  Turner}}]{klotsa05}
\bibinfo{author}{\bibfnamefont{D.}~\bibnamefont{Klotsa}},
  \bibinfo{author}{\bibfnamefont{R.~A.} \bibnamefont{Roemer}},
  \bibnamefont{and} \bibinfo{author}{\bibfnamefont{M.~S.}
  \bibnamefont{Turner}}, \bibinfo{journal}{Biophys. J.}
  \textbf{\bibinfo{volume}{89}}, \bibinfo{pages}{2187} (\bibinfo{year}{2005}).

\bibitem[{\citenamefont{Caetano and Schulz}(2005)}]{caetano04}
\bibinfo{author}{\bibfnamefont{R.~A.} \bibnamefont{Caetano}} \bibnamefont{and}
  \bibinfo{author}{\bibfnamefont{P.~A.} \bibnamefont{Schulz}},
  \bibinfo{journal}{Phys.\ Rev.\ Lett.} \textbf{\bibinfo{volume}{95}},
  \bibinfo{pages}{126601} (\bibinfo{year}{2005}).

\bibitem[{\citenamefont{Wang and Chakraborty}(2006)}]{tapash06}
\bibinfo{author}{\bibfnamefont{X.~F.} \bibnamefont{Wang}} \bibnamefont{and}
  \bibinfo{author}{\bibfnamefont{T.} \bibnamefont{Chakraborty}},
 \bibinfo{journal}{\texttt{cond-mat/0603672}}  (\bibinfo{year}{2006}).


\bibitem[{\citenamefont{Voityuk et~al.}(2000)\citenamefont{Voityuk, R{\"o}sch,
  Bixon, and Jortner}}]{bixon00}
\bibinfo{author}{\bibfnamefont{A.}~\bibnamefont{Voityuk}},
  \bibinfo{author}{\bibfnamefont{N.}~\bibnamefont{R{\"o}sch}},
  \bibinfo{author}{\bibfnamefont{M.}~\bibnamefont{Bixon}}, \bibnamefont{and}
  \bibinfo{author}{\bibfnamefont{J.}~\bibnamefont{Jortner}},
  \bibinfo{journal}{J.\ Phys.\ Chem.\ B} \textbf{\bibinfo{volume}{104}},
  \bibinfo{pages}{5661} (\bibinfo{year}{2000}).


\bibitem[{\citenamefont{Voityuk et~al.}(2001)\citenamefont{Voityuk, Siriwong, and R{\"o}sch}}]{voityuk01}
\bibinfo{author}{\bibfnamefont{A.}~\bibnamefont{Voityuk}},
 \bibinfo{author}{\bibfnamefont{K.}~\bibnamefont{Siriwong}}, \bibnamefont{and}
  \bibinfo{author}{\bibfnamefont{N.}~\bibnamefont{R{\"o}sch}},
  \bibinfo{journal}{Phys. Chem. Chem. Phys} \textbf{\bibinfo{volume}{3}},
  \bibinfo{pages}{5421} (\bibinfo{year}{2001}).


\bibitem[{\citenamefont{Grozema et al.}(2002)\citenamefont{Grozema, Siebbeles, Berlin, and Ratner}}]{berlin02}
\bibinfo{author}{\bibfnamefont{F.~C.}~\bibnamefont{Grozema}},
 \bibinfo{author}{\bibfnamefont{L.~D.~A.}~\bibnamefont{Siebbeles}}, 
  \bibinfo{author}{\bibfnamefont{Y.~A.}~\bibnamefont{Berlin}},\bibnamefont{and}
  \bibinfo{author}{\bibfnamefont{M.~A.}~\bibnamefont{Ratner}},
  \bibinfo{journal}{ChemPhysChem} \textbf{\bibinfo{volume}{6}},
  \bibinfo{pages}{536} (\bibinfo{year}{2002}).


\bibitem[{\citenamefont{Berlin et~al.}(2001)\citenamefont{Berlin, Burin, and Ratner}}]{berlin01}
\bibinfo{author}{\bibfnamefont{Y.~A.}~\bibnamefont{Berlin}},
 \bibinfo{author}{\bibfnamefont{A.~L.}~\bibnamefont{Burin}}, \bibnamefont{and}
  \bibinfo{author}{\bibfnamefont{M.~A.}~\bibnamefont{Ratner}},
  \bibinfo{journal}{J. Am. Chem. Soc.} \textbf{\bibinfo{volume}{123}},
  \bibinfo{pages}{260} (\bibinfo{year}{2001}).

\bibitem[{\citenamefont{Grozema et~al.}(2001)\citenamefont{Grozema, Berlin, and Siebbeles}}]{grozema00}
\bibinfo{author}{\bibfnamefont{F.~C.}~\bibnamefont{Grozema}},
 \bibinfo{author}{\bibfnamefont{Y.~A.}~\bibnamefont{Berlin}}, \bibnamefont{and}
 \bibinfo{author}{\bibfnamefont{L.~D.~A.}~\bibnamefont{Siebbeles}},
  \bibinfo{journal}{J. Am. Chem. Soc.} \textbf{\bibinfo{volume}{122}},
  \bibinfo{pages}{10903} (\bibinfo{year}{2000}).

 \bibitem[{\citenamefont{Palmero et~al.}(2004)\citenamefont{Palmero, Archilla, Hennig, and Romero}}]{hennig04}
 \bibinfo{author}{\bibfnamefont{F.}~\bibnamefont{Palmero}},
  \bibinfo{author}{\bibfnamefont{J.~F.~R.}~\bibnamefont{Archilla}},
 \bibinfo{author}{\bibfnamefont{D.}~\bibnamefont{Hennig}}, \bibnamefont{and}
 \bibinfo{author}{\bibfnamefont{F.~R.}~\bibnamefont{Romero}},
   \bibinfo{journal}{New J. Phys.} \textbf{\bibinfo{volume}{6}},
   \bibinfo{pages}{13} (\bibinfo{year}{2004}).


 \bibitem[{\citenamefont{ Senthilkumar et~al.}(2004)\citenamefont{Senthilkumar, Grozema, Fonseca, Bickelhaupt, Lewis, Berlin, Ratner, and Siebbeles}}]{siebbeles05}
 \bibinfo{author}{\bibfnamefont{K.}~\bibnamefont{Senthilkumar}},
  \bibinfo{author}{\bibfnamefont{F.~C.}~\bibnamefont{Grozema}},
 \bibinfo{author}{\bibfnamefont{C.}~\bibnamefont{Fonseca}},
 \bibinfo{author}{\bibfnamefont{F.~M.}~\bibnamefont{Bickelhaupt}},
 \bibinfo{author}{\bibfnamefont{F.~D.}~\bibnamefont{Lewis}},
 \bibinfo{author}{\bibfnamefont{Y.~A.}~\bibnamefont{Berlin}},
 \bibinfo{author}{\bibfnamefont{M.~A.}~\bibnamefont{Ratner}}, \bibnamefont{and}
 \bibinfo{author}{\bibfnamefont{L.~D.~A.}~\bibnamefont{Siebbeles}},
   \bibinfo{journal}{J. Am. Chem. Soc.} \textbf{\bibinfo{volume}{127}},
   \bibinfo{pages}{14894} (\bibinfo{year}{2005}).





\bibitem[{\citenamefont{Mahan}(2000)}]{mahan}
\bibinfo{author}{\bibfnamefont{G.~D.} \bibnamefont{Mahan}},
  \emph{\bibinfo{title}{Many-Particle Physics}} (\bibinfo{publisher}{Plenum
  Press}, \bibinfo{address}{New York}, \bibinfo{year}{2000}),
  \bibinfo{edition}{3rd} ed. 

\bibitem[{\citenamefont{Hewson and Newns}(1979)}]{hewson79}
\bibinfo{author}{\bibfnamefont{A.~C.} \bibnamefont{Hewson}} \bibnamefont{and}
  \bibinfo{author}{\bibfnamefont{D.~M.} \bibnamefont{Newns}},
  \bibinfo{journal}{J. Phys. C: Solid State Phys.}
  \textbf{\bibinfo{volume}{12}}, \bibinfo{pages}{1665} (\bibinfo{year}{1979}).

\bibitem[{\citenamefont{Meir and Wingreen}(1992)}]{mw92}
\bibinfo{author}{\bibfnamefont{Y.}~\bibnamefont{Meir}} \bibnamefont{and}
  \bibinfo{author}{\bibfnamefont{N.~S.} \bibnamefont{Wingreen}},
  \bibinfo{journal}{Phys.\ Rev.\ Lett.} \textbf{\bibinfo{volume}{68}},
  \bibinfo{pages}{2512} (\bibinfo{year}{1992}).

\bibitem[{\citenamefont{Felice et~al.}(2002)\citenamefont{Felice, Calzolari,
  and Molinari}}]{felice02}
\bibinfo{author}{\bibfnamefont{R.~Di} \bibnamefont{Felice}},
  \bibinfo{author}{\bibfnamefont{A.}~\bibnamefont{Calzolari}},
  \bibinfo{author}{\bibfnamefont{E.}~\bibnamefont{Molinari}},
\bibnamefont{and} \bibinfo{author}{\bibfnamefont{A.}~\bibnamefont{Garbesi}},
  \bibinfo{journal}{Phys.\ Rev.\ B} \textbf{\bibinfo{volume}{65}},
  \bibinfo{pages}{045104} (\bibinfo{year}{2002}).

\bibitem[{\citenamefont{Tran et~al.}(2000)\citenamefont{Tran, Alavi, and Gruner}}]{tran00}
\bibinfo{author}{\bibfnamefont{P.} \bibnamefont{Tran}},
  \bibinfo{author}{\bibfnamefont{B.}~\bibnamefont{Alavi}},
\bibnamefont{and} \bibinfo{author}{\bibfnamefont{G.}~\bibnamefont{Gruner}},
  \bibinfo{journal}{Phys.\ Rev.\ Lett.} \textbf{\bibinfo{volume}{85}},
  \bibinfo{pages}{1564} (\bibinfo{year}{2000}).


\bibitem[{\citenamefont{Rakitin et~al.}(2001)\citenamefont{Rakitin, Aich, Papadopoulos, Kobzar,  Vedeneev, Lee, and Xu}}]{rakitin01}
\bibinfo{author}{\bibfnamefont{A.} \bibnamefont{Rakitin}},
  \bibinfo{author}{\bibfnamefont{P.}~\bibnamefont{Aich}},
\bibinfo{author}{\bibfnamefont{C.}~\bibnamefont{Papadopoulos}},
\bibinfo{author}{\bibfnamefont{Y.}~\bibnamefont{Kobzar}},
\bibinfo{author}{\bibfnamefont{A.~S}~\bibnamefont{Vedeneev}},
\bibinfo{author}{\bibfnamefont{J.~S..}~\bibnamefont{Lee}},
\bibnamefont{and} 
\bibinfo{author}{\bibfnamefont{J.~M.}~\bibnamefont{Xu}},
  \bibinfo{journal}{Phys.\ Rev.\ Lett.} \textbf{\bibinfo{volume}{86}},
  \bibinfo{pages}{3670} (\bibinfo{year}{2001}).



\bibitem[{\citenamefont{Zalinge et al.~}(2006)\citenamefont{Zalinge, Schiffrin, Bates, Haiss, Ulstrup, and Nichols}}]{zalinge06}
\bibinfo{author}{\bibfnamefont{H.}~\bibnamefont{van Zalinge}},
  \bibinfo{author}{\bibfnamefont{D. J.}~\bibnamefont{Schiffrin}},
 \bibinfo{author}{\bibfnamefont{A. D.}~\bibnamefont{Bates}},
 \bibinfo{author}{\bibfnamefont{W.}~\bibnamefont{Haiss}},
 \bibinfo{author}{\bibfnamefont{J.}~\bibnamefont{Ulstrup}},
  \bibnamefont{and} \bibinfo{author}{\bibfnamefont{R. J.}~\bibnamefont{Nichols}},
  \bibinfo{journal}{ChemPhysChem} \textbf{\bibinfo{volume}{7}},
  \bibinfo{pages}{94} (\bibinfo{year}{2006}).




\end{thebibliography}

\end{document}